\documentclass[11pt, twosided]{article}

\usepackage{amsfonts}
\usepackage{amsmath}
\usepackage{graphicx}
\usepackage{color}

\begin{document}

\vskip 0.2cm
{\centerline{\Large {\bf Complex Mass Shells for Coloured quarks}}}
\vskip 0.2cm
{\centerline{\Large {\bf and their Asymptotic Confinement}}}
\vskip 0.4cm
\indent
\centerline{{\bf Richard Kerner$^a$ and Jerzy Lukierski$^b$}}
\vskip 0.3cm
\indent 
{\small a: Laboratoire de Physique Th\'eorique de la Mati\`ere}  \\
{\small Condens\'ee - CNRS URA 7600 - Sorbonne-Universit\'e, B.C. 121,} 
 \\
{\small 4 Place Jussieu 75005 Paris, France}\footnote{{\small e-mail  a: richard.kerner@sorbonne-universite.fr}
\vskip 0.1cm
{\small e-mail  b: jerzy.lukierski@uwr.edu.pl} }
\\ \\ 
{\small b: Institute of Theoretical Physics, Wroc{\l}aw University, }
 \\ 
{\small Plac Maxa Borna 9,  Wroc{\l}aw, Poland }
 \\

\begin{abstract}
{The present paper is the continuation of our previous work (\cite{RKJL2021}) where we introduced a $Z_3$-symmetric
covering of the Lorentz group as a natural symmetry describing the quark fields. 

In the current version of QCD quarks are described by coloured triplets of standard Dirac fields. In contrast,
we proposed to describe the colour triplets of quarks by entangled $Z_3$-graded Lee-Wick type fields,
one with real mass and the two remaining ones with mutually conjugate complex masses. This is obtained
by attributing colour degrees of freedom to six Pauli spinors, three endowed with colours and three with anti-colours,
which are united into one $12$-component generalized ``coloured Dirac spinor". 

Thus entangled triplet of quark fields is described on-shell by a linear Sch\"odinger-like system akin to the Dirac equation.
The sixth-order dispersion relations lead to solutions suitably vanishing in asymptotic region, exhibiting the well
established confinement property of coloured quarks' degrees of freedom. 

We add that in the so modified approach to QCD one should employ in the quark sector the $Z_3$-graded extension
of the Lorentz symmetries, which do not commute with hidden $SU(3)$ colour transformations. (see \cite{RKJL2021}, \cite{RK2018}).
Propagators and interaction with gluon and electromagnetic fields are discussed in the last section.}

\end{abstract}

\vskip 0.4cm
\section{Introduction}

In past few decades the $Z_3$ symmetry was studied with growing interest, and has led to generalizations
not only of the usual $Z_2$-graded algebraic structures, like Grassmann and Clifford algebras
(see \cite{Kerner1991}, \cite{Kerner2001}), but also of
less familiar {\it ternary algebras} (see \cite{VainKer}, \cite{Bazunova}, \cite{VARKOL}). 

This article is conceived as a continuation of our previous publication \cite{RKJL2021} in which a $Z_3$-symmetric generalization of the Lorentz symmetry
was proposed as the appropriate algebraic framework unifying in a non-trivial way the $SU(3)$-color algebra with the Lorentz algebra, resulting
in the colour Dirac equation for quarks. The most important result was the discovery that in order to close the $Z_3$-graded colour Lorentz algebra,
the twelve dimensional representation does not suffice, and one is forced to enlarge it to $72$ dimensions, which come from the tensor product
of the $12$-parameter space representing the degrees of freedom of three coloured Dirac spinors, by extra $6$ dimensions, corresponding to two flavors
 and three families. 

Having elucidated the representation properties of the coloured Dirac equation for quarks in our previous paper, we now shall analyze the properties
of its solutions. The most important feature of which is the pertaining presence of exponentially damped propagationn which might be interpreted
as algebraic confinement. We will also show how certain binary or ternary combinations of damped solutions can produce free running waves. 

In standard QCD setting the triplet of coloured quark fields is described by three independent Dirac spinors. However, one should take into consideration 
the property that due to the quark confinement mechanism the states with non vanishing colour charges, and in particular single quark states, were
never observed experimentally (\cite{Greensite}). One can deduce that such a feature leads to certain freedom in choosing how the free coloured quark 
fields may be defined, opening new vistas for investigation. In particular, one can look for modification of dynamical equations for colour quark fields 
leading to confinement properties obtained in algebraic way. {\footnote{In the standard quark model confinement is obtained from particular
dynamics imposed by the choice of $q {\bar{q}}$ and $qqq$ potentials.}}

Since the establishment by Han and Nambu \cite{HanNambu},  of the $SU(3)_{color}$ group as the fundamental symmetry 
of strong interactions between quarks carried by gluons, (\cite{Georgi}, \cite{Gross}, \cite{Politzer})

In a series of recently published papers (\cite{RKOS2012} - \cite{RKJL2021}) we studied the proposal consisting in entangling six Pauli spinors
with three colours and anti-colours into a system of $6$ two-component equations, thus producing the $Z_3$-graded modification of
the $12$-dimensional standard colour triplet of Dirac fields; symbolically, decomposing its $12$ degrees of freedom into $6 \times 2$
(the six Pauli spinors) instead of $3 \times 4$ (the three Dirac spinors). In such a way we obtained a $Z_3$-covariant triplet of quark fields
which satisfies the so-called {\it colour Dirac equation}. The choice of discrete cyclic $Z_3$ group of three elements, 
$(1, \; {\hat{a}}, \; {\hat{a}}^2)$, with ${\hat{a}}^3 = 1$ is linked with the choice of $SU(3)$ group for description of colour
symmetries. The fundamental one-dimensional complex representation is obtained if we choose ${\hat{a}} = j = e^{\frac{2 \pi i}{3}}$.


The $12$-dimensional colour Dirac multiplet is constructed out of six $2$-component Pauli spinors $\varphi_{\pm}, \; \chi_{\pm}, \; \psi_{\pm}$
representing three colours and three anti-colours, covariant under $Z_3 \times Z_2$ discrete symmetries.{\footnote{The colour Dirac equations
satisfied by this multiplet of Pauli spinors interpreted as ``colour Dirac spinor'', were introduced in (\cite{RKOS2013}) and (\cite{Kerner2018}).}}
The system of equations generalizing Dirac's equation (in its Schr\"odinger-like version) is as follows:
{\footnote{The system (\ref{systemsix}) can be given a relativistic form $ (\Gamma^{\mu} p_{\mu} - mc) \; \Psi = 0$ by multiplication 
from the lest by an appropriate $12 \times 12$ matrix. Then the mass term becomes proportional to the unit matrix, while the energy values
$E = c p_0$ display the spectrum characterizing the $Z_3 \times Z_2$ multiplet $\pm E, \; \pm jE$ and $\pm j^2 E$.}}

\begin{equation}
\begin{split}
& E \; \varphi_{+} = mc^2 \, \varphi_{+} + c \; {\boldsymbol{\sigma}} \cdot {\bf p} \, \chi_{-},
\\
&E \; \varphi_{-} = - mc^2 \, \chi_{-} + c \; {\boldsymbol{\sigma}} \cdot {\bf p} \, \chi_{+},
\\
& E \; \chi_{+} = j mc^2 \, \chi_{+} + c \; {\boldsymbol{\sigma}} \cdot {\bf p} \, \psi_{-},
\\
& E \; \chi_{-} = - j mc^2 \, \chi_{-} + c \; {\boldsymbol{\sigma}} \cdot {\bf p} \, \psi_{+}
\\
& E \; \psi_{+} =j^2  mc^2 \, \psi_{+} + c \; {\boldsymbol{\sigma}} \cdot {\bf p} \, \varphi_{-},
\\
&E \; \psi_{-} = - j^2 \;mc^2 \, \psi_{-} + c \; {\boldsymbol{\sigma}} \cdot {\bf p} \, \varphi_{+},
\end{split}
\label{systemsix}
\end{equation}
The  system (\ref{systemsix}) is invariant under the following discrete symmetries:
\vskip 0.2cm
\indent
- the particle-antiparticle symmetry $Z_2$ group realized as $(1, {\hat{\tau}})$, where ${\hat{\tau}}$ denotes involution $({\hat{\tau}}^2 = 1)$
and is represented on the set (\ref{systemsix}) by the involution
\begin{equation}
{\hat{\tau}}: \; \; \; m \rightarrow -m, \; \; \; ( \varphi_{\pm}, \; \chi_{\pm}, \; \psi_{\pm} ) \rightarrow ( \varphi_{\mp}, \; \chi_{\mp}, \; \psi_{\mp})
\label{papsym}
\end{equation}
\vskip 0.2cm
\indent
- the colour mixing symmetry $Z_3$ is realized by the following simultaneous maps:
$$
{\hat{a}}: \; \; \; m \rightarrow j \; m, \; \; \; \;  \varphi_{\pm} \rightarrow \chi_{\pm} \rightarrow \psi_{\pm} \rightarrow \varphi_{\pm},$$
\begin{equation}
{\hat{a}}^2:  \; \; \; m \rightarrow j^2 \; m, \; \; \; \;  \varphi_{\pm} \rightarrow \psi_{\pm} \rightarrow \chi_{\pm} \rightarrow \varphi_{\pm},
\label{clorsym}
\end{equation}
One can show that each of the $12$ components of the ``colour Dirac spinor'', $\varphi^1_{+}, \; \varphi^2_{+}, \; \varphi^1_{-}, \; \varphi^2_{-},$
etc., satisfy the following sixth-order equation:
\begin{equation}
\left( E^6 -c^6 {\bf p}^6 \right) \; \Psi = m^6 c^{12} \; \Psi.
\label{KGsixth}
\end{equation} 
where $\Psi= (\varphi_{+}, \; \varphi_{-}, \; \chi_{+}, \; \chi_{-}, \; \psi_{+}, \; \psi_{-} )^T$

It should be stressed here that the system (\ref{systemsix}) represents the same number of linear equations (i.e. $12$) and the same number of field variables as
the system of three Dirac equations for three Dirac spinors, only reshuffled in a different way, displaying an additional $Z_3$ symmetry.

According to the correspondence principle between classical and quantum-mechanical observables, in Schr\"odinger picture, 
the energy and momentum are represented by the following differential operators:
\begin{equation}
E \rightarrow - i \hbar \frac{\partial}{\partial t}, \; \; \; {\bf p} \rightarrow - i \hbar {\boldsymbol{\nabla}}.
\label{Epcorresp}
\end{equation}
Then the differential equation corresponding to (\ref{KGsixth}) is an analogue of the Klein-Gordon equation (with $\hbar$ and $c$ put equal to $1$):
\begin{equation}
\left[ \partial^2_t - \Delta - m^2 \right] \Psi_k = 0 \; \rightarrow \; \left[ \partial^6_t - {\Delta}^3 - m^6 \right] \Psi_k = 0
\label{dalembert}
\end{equation} 
From (\ref{dalembert}) we obtain, by analogy with the Klein-Gordon equation,  the following colour generalization of the relativistic 
dispersion rule, relating energy, momentum and mass:
\begin{equation}
\omega = \sqrt{{\bf p}^2 + m^2 c^2 }, \; \rightarrow \; \Omega = \root6\of{ \mid {\bf p} \mid^6 + m^6},
\label{BigOmega}
\end{equation}
A more rigogous derivation of these dispersion relations is given in the next section, where they are obtained as determinants of the operators 
in matrix representations.

\section{Matrix representation}

It was shown in previously published papers that the system (\ref{systemsix}) can be encoded into what we can call the ``colour Dirac equation''
using generalized Dirac matrices and the relativistic $4$-vector notation.
Let us recall the standard form of the Dirac equation written by means of the first-order matrix-valued differential oprator, acting on the
 four-component Dirac spinors:
\begin{equation}
\gamma^{\mu} p_{\mu} \; \psi = mc \; {\mbox{l\hspace{-0.55em}1}}_4 \; \psi, \; \; \; \mu, \nu,.. = 0,1,2,3
\label{Diracgamma}
\end{equation}
where $p_0 = E/c$ and the four Dirac matrices are defined by means of the $2 \times 2$ hermitian Pauli matrices as follows: 
$$\gamma^0 = \sigma_3 \otimes {\mbox{l\hspace{-0.55em}1}}_2, \; \; \; \gamma^k = (i \sigma_2) \otimes \sigma^k.$$
The structure of the tensor products reflects the $Z_2 \times Z_2$ symmetry: the first one corresponding to the half-integer spin of the Dirac particle
(which was the electron, in its first intention) and the charge conjugation, corresponding to the particle-antiparticle symmetry.

It is also worthwhile to observe that taking the determinant of both sides of the system (\ref{Diracgamma}) leads to the $4$-th order 
characteristic equation 
\begin{equation}
{\rm det} \left( \gamma^{\mu} p_{\mu} \right) = (\frac{E^4}{c^4} - \mid {\bf p} \mid^4) = {\rm det}  (mc \; {\mbox{l\hspace{-0.55em}1}}_4 ) = m^4 c^4,
\label{detgamma}
\end{equation}
which displays double degeneracy, being the equality of ful squares, so that the effective characteristic equation is $E^2 - \mid {\bf p} \mid^2 c^2 = m^2 c^4.$

The system (\ref{systemsix}) of twelve equations intertwining six Pauli spinors can be also represented as a $12 \times 12$ matrix valued operator, linear in
energy, momentum and mass. Instead of a tensor product the usual Dirac operator with the $3 \times 3$ unit matrix, ${\mbox{l\hspace{-0.55em}1}}_3$, the $Z_3 \times Z_2 \times Z_2$
symmetry introducing the three colour degrees of freedom, we have here the following matrix structure:
\begin{equation}
E \; {\mbox{l\hspace{-0.55em}1}}_3 \otimes {\mbox{l\hspace{-0.55em}1}}_2 \otimes {\mbox{l\hspace{-0.55em}1}}_2 \; \Psi =
Q_3 \otimes \sigma_1 \otimes {\boldsymbol{\sigma}} \cdot {\bf p} \; \Psi + mc^2 \; B \otimes \sigma_3 \otimes {\mbox{l\hspace{-0.55em}1}}_2 \; \Psi.
\label{Ediag}
\end{equation}                                                                                                                                  

where $B$ and $Q_3$ are the following $3 \times 3$ matrices:
{\small \begin{equation}
B = \begin{pmatrix} 1 & 0 & 0 \cr 0 & j & 0 \cr 0 & 0 & j^2 \end{pmatrix} \; \; {\rm and} \; \; 
Q_3 = \begin{pmatrix} 0 & 1 & 0 \cr 0 & 0 & 1 \cr 1 & 0 & 0 \end{pmatrix}.
\label{BQmatrices}
\end{equation} }
Their products and powers generate the following set of eight traceless $3 \times 3$ matrices (plus the unit $3 \times 3$ matrix ${\mbox{l\hspace{-0.55em}1}}_{3}$):
{\small \begin{equation}
Q_1 = \begin{pmatrix}  0 & 1 & 0 \cr 0 & 0 & j \cr j^2 & 0 & 0 \end{pmatrix}, \; 
Q_2 = \begin{pmatrix}  0 & 1 & 0 \cr 0 & 0 & j^2 \cr j & 0 & 0 \end{pmatrix}, \; 
Q_3 = \begin{pmatrix}  0 & 1 & 0 \cr 0 & 0 & 1 \cr 1 & 0 & 0 \end{pmatrix}, \;  
\label{threeQ1}
\end{equation} 
\begin{equation}
Q^{\dagger}_1 = \begin{pmatrix}  0 & 0 & j \cr 1 & 0 & 0 \cr 0 & j^2 & 0 \end{pmatrix}, \; 
Q^{\dagger}_2 = \begin{pmatrix}  0 & 0 & j^2 \cr 1 & 0 & 0 \cr 0 & j & 0 \end{pmatrix}, \; 
Q^{\dagger}_3 = \begin{pmatrix}  0 & 0 & 1 \cr 1 & 0 & 0 \cr 0 & 1 & 0 \end{pmatrix}, \; 
\label{threeQdagger1}
\end{equation} } 
where $j$ is the third primitive root of unity, 
\begin{equation} j = e^{\frac{2 \pi i}{3}}, \; \; j^2 = e^{\frac{4 \pi i}{3}}, \; \; 1+j+j^2 = 0.
\label{Jot1}
\end{equation}
and the following two linearly independent {\it diagonal} matrices: 
\begin{equation}
B = \begin{pmatrix}  1 & 0 & 0 \cr 0 & j & 0 \cr 0 & 0 & j^2 \end{pmatrix}, \; \; \; \; B^{\dagger} = 
\begin{pmatrix}  1 & 0 & 0 \cr 0 & j^2 & 0 \cr 0 & 0 & j \end{pmatrix}.
\label{twoBmatrices1}
\end{equation}
${\cal{M}}^{\dagger}$ denotes the Hermitean conjugate of matrix ${\cal{M}}$. We see that all matrices $Q; \;Q^{\dagger}, B$ and $B^{\dagger}$ 
are non-Hermitean, but they are mutually Hermitean conjugates. They span a basis for the $SU(3)$ algebra alternative with respect to the
well-known Gell-Mann matrices $\lambda_k, \; k = 1, 2,...8.${\footnote{For the first time such a representation of $SU(3)$ Lie algebra was introduced by
V.G. Kac (\cite{Kac1994}); see also the Appendix I.}}
Moreover, we have 
\begin{equation}
Q_a^2 = Q_a^{\dagger} = Q_a^{-1}, \; \; B^{\dagger} = B^{-1},
\label{QBSU3}
\end{equation}
so that all these matrices satisfy the relation defining the $SU(3)$ group, which is the collection of all unitary $3 \times 3$ matrices satisfying 
the extra condition $M^{\dagger} = M^{-1}$.
Therefore, all the eight matrices form a finite subset of the Lie group $SU(3)$, but it is not a subgroup, even if we add the unite $3 \times 3$ matrix.
   
The multiplication table, given in the Appendix I.   

The relation of this $SU(3)$ matrix basis with the currently used Gell-Mann matrices is given in the Appendix I.

Moreover, the automorphisms of this set conserving property define the entire $SU(3)$ group. The proof is as follows:
let $M^{\dagger} = M^{-1}$ Consider the similarity transformation ${\tilde{M}} = S M S^{-1}$, and require the same property
for ${\tilde{M}}$, i.e. ${\tilde{M}}^{\dagger} = {\tilde{M}}^{-1}$; But:
\begin{equation}
{\tilde{M}}^{\dagger} = \left( S M S^{-1} \right)^{\dagger} = (S^{-1})^{\dagger} M^{\dagger} S^{\dagger} = (S^{-1})^{\dagger} M^{-1} S^{\dagger}.
\label{SU3prop1}
\end{equation}
On the other hand, we have
\begin{equation}
{\tilde{M}}^{-1} = ( S M S^{-1})^{-1} = S M^{-1} S^{-1}.
\label{SU3prop2}
\end{equation}
Comp)aring the two relations, we see that one must have $S^{\dagger} = S^{-1}$; therefore, the group of automorphisms of the finite
subgroup generated by the nonions is the entire group $SU(3)$ in its fundamental representation.

Therefore, all these matrices form a finite subset of the Lie group $SU(3)$ - as a matter of fact 
Now, if we want to get as close as possible to the relativistic form of Dirac's equation given in (\ref{Diracgamma}), we should multiply all matrix operators
by the inverse of the $12 \times 12$ matrix standing before the mass term $mc^2$. The inverse is given by the following tensor product of matrices:
\begin{equation}
\left( B \otimes \sigma_3 \otimes {\mbox{l\hspace{-0.55em}1}}_2 \right)^{-1} = B^{\dagger} \otimes \sigma_3 \otimes {\mbox{l\hspace{-0.55em}1}}_2,
\label{Binverse}
\end{equation}
with as usual, $ {\mbox{l\hspace{-0.55em}1}}_2 = \begin{pmatrix} 1 & 0 \cr 0 & 1 \end{pmatrix}, \; \; \; 
\sigma_1 = \begin{pmatrix} 0 & 1 \cr 1 & 0 \end{pmatrix}, \; \; \; 
\sigma_3 = \begin{pmatrix} 1 & 0 \cr 0 & -1 \end{pmatrix}. $  
Multiplying the system (\ref{Ediag}) by $B^{\dagger} \otimes \sigma_3 \otimes {\mbox{l\hspace{-0.55em}1}}_2$ from the left and dividing by $c$, we get the new form
of our equation,
\begin{equation}
\frac{E}{c} \;  B^{\dagger} \otimes \sigma_3 \otimes {\mbox{l\hspace{-0.55em}1}}_2
- Q_2 \otimes (i \sigma_2) \otimes  {\boldsymbol{\sigma}}\cdot {\bf p} = 
m c \;  {\mbox{l\hspace{-0.55em}1}}_3 \otimes {\mbox{l\hspace{-0.55em}1}}_2 \otimes {\mbox{l\hspace{-0.55em}1}}_2,
\label{Gammafirst}
\end{equation}
where we used the fact that under matrix multiplication,
$\sigma_3 \sigma^3 = {\mbox{l\hspace{-0.55em}1}}_2$,
$B^{\dagger} B = {\mbox{l\hspace{-0.55em}1}}_3$ and $B^{\dagger} Q_3 =  Q_2$.
with 
$$Q_2 = \begin{pmatrix} 0 & 1 & 0 \cr 0 & 0 & j^2 \cr j & 0 & 0 \end{pmatrix}. $$
The equation (\ref{Gammafirst}) can be written in a concise manner using the Minkowskian indices 
and the usual pseudo-scalar product of two four-vectors as follows:
\begin{equation}
\Gamma^{\mu} p_{\mu} \; \Psi = m c \; {\mbox{l\hspace{-0.55em}1}}_{12} \; \Psi, \; \; \; 
{\rm with} \; \; p^0 = \frac{E}{c}, \; \; p^k = [ \; p^x, p^y, p^z \; ].
\label{Gammasecond}
\end{equation}
with $12 \times 12$ matrices $\Gamma^{\mu}, \; \; (\mu = 0, 1, 2, 3)$ defined as follows:
\begin{equation}
\Gamma^0 = \sigma_3 \otimes B^{\dagger} \otimes {\mbox{l\hspace{-0.55em}1}}_2, \; \; \; \; \; 
\Gamma^{k} =  i \sigma_2 \otimes  Q_2 \otimes  {\sigma}^k
\label{Gammamu}
\end{equation}

Now the system (\ref{systemsix}) can be represented in a Dirac-like form as follows:
\begin{equation}
\Gamma^{\mu} p_{\mu} \Psi = mc \; {\mbox{l\hspace{-0.55em}1}}_{12} \; \Psi,
\label{Terndirac0}
\end{equation} 
where $\Psi$ is the generalized $12$-component spinor made of  $6$ Pauli spinors, and the generalized
$12 \times 12$ Dirac matrices $\Gamma^{\mu}$ are constructed as follows:
\begin{equation}
\Gamma^0 = B^{\dagger} \otimes \sigma_3 \otimes {\mbox{l\hspace{-0.55em}1}}_{2}, \; \; \; 
\Gamma^i = Q_2 \otimes (i \sigma_2) \otimes \sigma^i,
\label{Gammasbig}
\end{equation}
Taking the determinant on both sides we get:
\begin{equation}
{\rm det } \left( \Gamma^{\mu} p_{\mu} \right) = (p_0^6 - \mid {\bf p} \mid^6)^2 = m^{12} c^{12},
\label{detDir12}
\end{equation}
which is {\it the square} of the characteristic equation $E^6/c^6 - \mid {\bf p } \mid^6 = m^6 c^6,$
or in terms of Fourier transforms, $k_0^6 - \mid {\bf k} \mid^6 = m^6.$

It is worthwhile to remind the similar result in the case of the usual Dirac equation, when one evaluates the determinant of the Dirac operator
in Fourier representation:
\begin{equation}
 {\rm det} \left( \gamma^{\mu}p_{\mu} - m c \; {\mbox{l\hspace{-0.55em}1}}_{4} \right) = (p_0^2 - \mid {\bf p } \mid^2 - m^2 c^2)^2.
\label{Diracdet}
\end{equation}
Note that here, like in the case of Maxwell's equations in vacuo, the characteristic equation is of fourth order, but it is degenreate -
here is only one characteristic surface instead of two. 

Let us conclude this section by recalling that the sixth-order expression on the left-hand side of (\ref{detDir12}) can be decomposed
into a product of three second-order ones,
\begin{equation}
k_0^6 - \mid {\bf k} \mid^6 = (k_0^2 - \mid {\bf k} \mid^2)(k_0^2 - j \mid {\bf k} \mid^2)(k_0^2 - j^2 \mid {\bf k} \mid^2),
\label{decomposition3}
\end{equation}
of which the first multiplier is manifestly Lorentz-invariant, while the second and third one are invariant under generalized
complex representations of a $Z_3$-covering of the Lorentz group (see the detailed discussion in \cite{RKJL2021}) 

\section{The Lorentz invariance}
\vskip 0.3cm
\indent
Although the characteristic equation $p_{\mu} p^{\mu} = \left( \frac{E}{c} \right)^6 - \mid {\bf p} \mid^6 = m^6 c^6 $
where $p^{\mu} = [E/c, {\bf p}]$ is not invariant under the usual Lorentz transformation 
$$p^{\mu} \rightarrow p^{\mu'} = \Lambda^{\mu'}_{\mu} p^{\mu},$$
let us show that the $12 \times 12$ matrix operator conserves its eigenvalues
and characteristic equation under generalized Lorentz transformations tensorized with the $Z_3$ group. 

Let us remark, for the sake of simplicity, that the right-hand side is a constant, thus invariant by definition under the transformations 
we are looking for; this is why it will be sufficient to consider exclusively the transformation properties of the homogeneous equation 
 $\left( \frac{E}{c} \right)^6 - \mid {\bf p} \mid^6 = 0.$
Next, the problem can be always reduced to two dimensions by choosing a Cartesian coordinate system with its $x$-axis aligned on the vector ${\bf p}$. 
The we shall have $p^{\mu} = [p^0, p^x, 0, 0]$, so that it will be enough to consider the Lorentz boosts involving only two components, $p_0 = E/c$ and $p^x$.
With the choice of units such that $c=1$, denoting $p^x$ simply by $p$, the generalized Dirac operator takes on the following form:
\vskip 0.3cm
${\small
\left(\begin{array}{ c c c |c c c |c c c |c c c| c c c| c c}
E & 0 && 0 & 0 && 0 & 0 && 0 & 0 && 0 & p && 0 & 0 \\
0 & E && 0 & 0 && 0 & 0 && 0 & 0 && p & 0 && 0 & 0 \\ \hline
0 & 0 && j^{2}E & 0 && 0 & 0 && 0 & 0 && 0 & 0 && 0 & j^{2} p\\
0 & 0 && 0 & j^{2}E && 0 & 0 && 0 & 0 && 0 & 0 && j^{2} p & 0\\ \hline
0 & 0 && 0 & 0 && j E & 0 && 0 & jp && 0 & 0 && 0 & 0 \\
0 & 0 && 0 & 0 && 0 & j E && jp & 0 && 0 & 0 && 0 & 0 \\ \hline
0 & 0 && 0 & -p && 0 & 0 && -E & 0 && 0 & 0 && 0 & 0\\
0 & 0 && -p & 0 && 0 & 0 && 0 & -E && 0 & 0 && 0 & 0\\ \hline
0 & 0 && 0 & 0 && 0 & -j^{2}p && 0 & 0 && -j^{2}E & 0 && 0 & 0\\
0 & 0 && 0 & 0 && -j^{2}p & 0 && 0 & 0 && 0 & j^{2}E && 0 & 0\\ \hline
0 & -jp && 0 & 0 && 0 & 0 && 0 & 0 && 0 & 0 && -jE & 0 \\
-jp & 0 && 0 & 0 && 0 & 0 && 0 & 0 && 0 & 0 && 0 & -jE 
\end{array}
 \right)} $
\vskip 0.3cm
The determinant of this matrix is equal to:
\begin{equation}
\det(A) = (E^{6} - p^{6})^{2}
\end{equation} 
and it has six distinct eigenvalues:
$$
(E^{6} - p^{6})^{1/6}, \; \; \;  j (E^{6} - p^{6})^{1/6}, \; \; \; j^2 (E^{6} - p^{6})^{1/6}, $$
\begin{equation}
- (E^{6} - p^{6})^{1/6}, \; \; \; -j (E^{6} - p^{6})^{1/6}, \; \; -j^2 (E^{6} - p^{6})^{1/6}.
\label{eigensix}
\end{equation}

 Consider now another $12 \times 12$ matrix $B$, containing the same entries differently disposed:
\vskip 0.3cm
${\small
\left(\begin{array}{ c c c |c c c |c c c |c c c| c c c| c c}
0 & 0 && 0 & 0 && 0 & 0 && 0 & 0 && E & p && 0 & 0 \\
0 & 0 && 0 & 0 && 0 & 0 && 0 & 0 && p & E && 0 & 0 \\ \hline
0 & 0 && 0 & 0 && 0 & 0 && 0 & 0 && 0 & 0 && E & j p\\
0 & 0 && 0 &0 && 0 & 0 && 0 & 0 && 0 & 0 && j p & E\\ \hline
0 & 0 && 0 & 0 && 0 & 0 && E & j^{2}p && 0 & 0 && 0 & 0 \\
0 & 0 && 0 & 0 && 0 & 0 && j^{2}p & E && 0 & 0 && 0 & 0 \\ \hline
0 & 0 && E & -p && 0 & 0 && 0 & 0 && 0 & 0 && 0 & 0\\
0 & 0 && -p & E && 0 & 0 && 0 & 0 && 0 & 0 && 0 & 0\\ \hline
0 & 0 && 0 & 0 && E & -jp && 0 & 0 && 0 & 0 && 0 & 0\\
0 & 0 && 0 & 0 && -jp & E && 0 & 0 && 0 & 0 && 0 & 0\\ \hline
E & -j^{2}p && 0 & 0 && 0 & 0 && 0 & 0 && 0 & 0 && 0 & 0 \\
-j^{2}p & E && 0 & 0 && 0 & 0 && 0 & 0 && 0 & 0 && 0 & 0 
\end{array} 
 \right)} $ 
\vskip 0.2cm
{Credit: Prof. Katarzyna G{\`o}rska }

\vskip 0.3cm
It is easy to check that it has the same determinant $\det(B) = (E^{6} - p^{6})^{2}$  as the colour Dirac operator, and the same eigenvalues \ref{eigensix}.
$\det(B) = \det(A) =(E^{6} - p^{6})^{2}.$

Using the Fourier transform, we can replace $E$ by $k_0$ and ${\bf p}$ by ${\bf k}$. 
Then the characteristic equation can be split in three factors as follows:
$$
k_0^6 - \mid {\bf k} \mid^6 = \left( k_0^2 - \mid {\bf k} \mid^2 \right)  \left( k_0^2 - j \; \mid {\bf k} \mid^2 \right)
 \left( k_0^2 - j^2 \; \mid {\bf k} \mid^2 \right)=$$
$$ =\left( k_0^2 - \mid {\bf k} \mid^2 \right) 
\; \left( k_0^4 + k_0^2  \mid {\bf k} \mid^2 + \mid {\bf k} \mid^4 \right) = 0.$$
The first factor defines the usual light cone, while the factor of degree four is strictly positive besides the origin $0$. 
The system has only one characteristic surface which is the same for all massless fields.
Each of the three factors remains invariant under a different representation of the $SL (2, {\bf C})$ group.
Let us introduce three matrices representing the same four-vector $k^{\mu}$:
\begin{equation}
K_3 = \begin{pmatrix} k_0 & k_x \cr k_x & k_0 \end{pmatrix}, \; \; K_1 = \begin{pmatrix} k_0 & j  k_x \cr j k_x & k_0 \end{pmatrix},
\; \; K_2 = \begin{pmatrix} k_0 & j^2  k_x \cr j^2 k_x & k_0 \end{pmatrix},
\label{threekmat}
\end{equation}  
whose determinants are, respectively,
\begin{equation}
{\rm det} K_1 = k_0^2 - j^2 k_x^2, \; \; \; {\rm det} K_2 = k_0^2 - j k_x^2, \; \; {\rm det} K_3 = k_0^2 -  k_x^2.
\label{threedets}
\end{equation} 

Note that only the third matrix $K_3$ is hermitian, and corresponds to a {\it real} space-time vector $k^{\mu}$,
while neither of the remaining two matrices $K_1$ and $K_2$ is hermitian; however, one is the hermitian conjugate of another, $K_1^{\dagger} = K_2.$

In what follows, we shall replace the absolute value of the wave vector $\mid {\bf k} \mid$ by a single spatial component, say  $k_x$, 
because for any given $4$-vector $k^{\mu} = [k_0, {\bf k} ]$ we can choose the coordinate system aligning its $x$-axis along the vector ${\bf k}$. 
Now it is easy to check that one has: 
$$ \begin{pmatrix} \cosh u & \sinh u \cr \sinh u & \cosh u \end{pmatrix} \begin{pmatrix} k_0 \cr k_x \end{pmatrix} = 
\begin{pmatrix} {k'}_0 \cr {k'}_x \end{pmatrix}, \; \; \; \; \; 
\begin{pmatrix} \cosh u & j^2 \sinh u \cr j \sinh u & \cosh u \end{pmatrix} \begin{pmatrix} k_0 \cr j \; k_x \end{pmatrix} = 
\begin{pmatrix} {k'}_0 \cr j \; {k'}_x \end{pmatrix} $$ 
\begin{equation}
\begin{pmatrix} \cosh u & j \sinh u \cr j^2 \sinh u & \cosh u \end{pmatrix} \begin{pmatrix} k_0 \cr j^2 k_x \end{pmatrix} = 
\begin{pmatrix} {k'}_0 \cr j^2 {k'}_x \end{pmatrix} 
\end{equation}

The transformed vectors are given by the following expressions:
$$ \hskip 0.5cm i) \;  k^{'}_0 = k_0  \cosh u + k_x \; \sinh u, \; \; \; k^{'}_x =   k_0  \sinh u +  k_x \; \cosh u $$
$$\hskip 0.3cm  ii) \; k^{'}_0 = k_0  \cosh u + j^2 \; k_x \; \sinh u, \; \; \; k^{'}_x =  j \; k_0  \sinh u +  k_x \; \cosh u $$
$$iii) \; k^{'}_0 = k_0  \cosh u + j \; k_x \; \sinh u, \; \; \; k_x^{'} =  j^2 \; k_0  \sinh u +  k_x \; \cosh u $$
Let us now introduce a $6 \times 6$ matrix composed out of the above three $2 \times 2$ matrices: 
 \begin{equation}
\begin{pmatrix} 0 & k_0 \; {\mbox{l\hspace{-0.55em}1}}_{2} + {\bf k} \cdot {\boldsymbol{\sigma}} & 0 \cr 
 0 & 0 &  k_0 \; {\mbox{l\hspace{-0.55em}1}}_{2} + j \, {\bf k} \cdot {\boldsymbol{\sigma}}  \cr
k_0 \; {\mbox{l\hspace{-0.55em}1}}_{2}+ j^2 \, {\bf k} \cdot {\boldsymbol{\sigma}}& 0 & 0 \end{pmatrix}
\label{threekays}
\end{equation}
Under the action of Lorentz boost represented in three different ways on the corresponding three
$2 \times 2$ submatrices, the determinant of the $6 \times 6$ matrix remains invariant.
\vskip 0.2cm
The ternary Dirac operator is composed of two similar $6 \times 6$ matrices, and its determinant, which is the square of det $K$, 
remains invariant under this particular representations of the Lorentz group intertwined with the $Z_3$ group.
The spatial rotations around the axes $0x, \; 0y$ and $0z$ are represented in the usual $4$-dimensional Minkowskian space as follows:
\begin{equation}
J_x = \begin{pmatrix} 0 & 0 & 0 & 0 \cr 0 & 0 & 0 & 0 \cr 0 & 0 & 0 & -1 \cr 0 & 0 & 1 & 0 \end{pmatrix}, \; \;  
J_y = \begin{pmatrix} 0 & 0 & 0 & 0 \cr 0 & 0 & 0 & 1 \cr 0 & 0 & 0 & 0 \cr 0 & -1 & 0 & 0 \end{pmatrix}, \; \;  
J_z = \begin{pmatrix} 0 & 0 & 0 & 0 \cr 0 & 0 & -1 & 0 \cr 0 & 1 & 0 & 0 \cr 0 & 0 & 0 & 0 \end{pmatrix}. 
\label{ThreeJ}
\end{equation}
The full set of $12 \times 12$ matrices representing three independent spatial rotations acting on the
twelve-dimensional $Z_3$-graded Minkowskian spacetime is as follows:
\begin{equation}
{\overset{(0)}{\cal{J}}}_i = \begin{pmatrix} J_i & 0 & 0 \cr 0 & J_i & 0 \cr 0 & 0 & J_i \end{pmatrix}, \; \; \; 
{\overset{(1)}{\cal{J}}}_i = \begin{pmatrix} 0 & J_i & 0 \cr 0 & 0 & J_i \cr J_i & 0 & 0 \end{pmatrix}, \; \; \; 
{\overset{(2)}{\cal{J}}}_i = \begin{pmatrix} 0 & 0 & J_i \cr J_i & 0 & 0 \cr 0 & J_i & 0 \end{pmatrix}, \; \; \;  
\label{ThreespaceJ}
\end{equation}
They also form a $Z_3$ graded Lie algebra with respect to the ordinary Lie bracket (the commutator of matrices). The $Z_3$-grades of the operators 
${\overset{(0)}{\cal{J}}}_i, \; {\overset{(1)}{\cal{J}}}_i, \; \, {\overset{(2)}{\cal{J}}}_i$ are as indicated by the overscripts $0,1,2$, and add
up modulo $3$ under multiplication.

Therefore we get the full set of $Z_3$-graded relations defining the algebra ($r, s, \; r+s$ are modulo $3$), conformally with the structure
of the $Z_3$-graded Lorentz algebra introduced in \cite{RKJL2019}.
\begin{equation}
[ \; {\overset{(r) }{{\cal{K}}_i}}, \; {\overset{(s) }{{\cal{K}}_k}} \; ]  = - \epsilon_{ikl} {\overset{(r+s) }{{\cal{J}}_l}}, \; \; \; \; \; 
[ \; {\overset{(r) }{{\cal{J}}_i}}, \; {\overset{(s) }{{\cal{K}}_k}} \; ]  = \epsilon_{ikl} {\overset{(r+s) }{{\cal{K}}_l}}, \; \; \; \; \;%
[ \; {\overset{(r) }{{\cal{J}}_i}}, \; {\overset{(s) }{{\cal{J}}_k}} \; ] = \epsilon_{ikl} {\overset{(r+s) }{{\cal{J}}_l}}.
\label{modulocomm}
\end{equation}

\section{Solutions}

The sixth-order characteristic equation $E^6 = \mid {\bf p} \mid^6 c^6 = m^6 c^{12}$ yields the following sixth-order differential equation after applying
the quantum corespondence principle according to which 
$E \rightarrow - i \hbar \partial_t, \; \; p_k \rightarrow - i \hbar \partial_k, \; \; k = 1,2,3,$
\begin{equation}
- \hbar^6 \frac{\partial^6}{\partial t^6} = - \hbar^6 \Delta^6 c^6 + m^6 c^{12}.
\label{diffeq6}
\end{equation} 
Any system of linear equations with constant coefficients has solutions expressible in terms of exponential functions :
\begin{equation}
f(t,x) \simeq C \; e^{\omega t - {\bf k} \cdot {\bf r}}, \; \; \; {\rm with} \; \; \; {\bf k} \cdot {\bf r} = k_x x + k_y y + k_z z.
\label{expsolution}
\end{equation}
In a given Cartesian reference frame it is possible to align the $Ox$ axis along the $3$-dimensional wave vector ${\bf k}$, so that
the scalar product ${\bf k} \cdot {\bf r}$ reduces to $k_x x$, which we shall denote simply by $kx$ from now on, without risk of confusion.

The differential equation is also reduced to the simplest form, involving only two partial derivatives, $\partial_t$ and $\partial_x$:
\begin{equation}
-\frac{1}{c^6} \; \frac{\partial^6 f}{\partial t^6} = \frac{\partial^6 f}{\partial x^6} - \mu^{6} f, \; \; \; {\rm with} \; \; \; \mu = \frac{m c}{\hbar}
\label{diffeq6bar}
\end{equation}  
Such a system is easily solved by the standard procedure of separation of variables: let us assume that the solution can be found as a product
of two functions, one depending on $t$, another on $x$:
\begin{equation}
f(t, x) = T(t) X(x).
\label{TX}
\end{equation}
Inserting into the equation (\ref{diffeq6bar}) and dividing by $XT$ we arrive at the following equation relating two ordinary sixth-order derivatives:
\begin{equation}
\frac{1}{T} \; \frac{1}{c^6} \; \frac{d^6 T}{dt^6} = \frac{1}{X} \; \frac{d^6 X}{dx^6} - \mu^6 = {\rm Constant} 
\label{separation}
\end{equation}

From a purely mathematical point of view, the sign of the constant, common to both expressions, is arbitrary - it can be negative, positive or zero.
Formally, we can put $Constant = \pm \frac{\omega^6}{c^6}$, without deciding about the constant $\omega$ itself: it can be real or imaginary, and also
multiplied by $j$ or $j^2$ without changing the sixth power. The exponential solutions are sought in the form of
\begin{equation}
T = e^{\omega t}, \; \; \; X = e^{kx}, \; \; \; {\rm provided \; that} \; \; \; \frac{\omega^6}{c^6} = k^6 - \mu^6.
\label{dispersopn1}
\end{equation}
Assuming that the mass parameter $\mu$ is either real or zero (massless quarks being not always excluded, especially according to certain pre-symmetry
breaking scenarios), the term $- \mu^6$ must be a negative real number. Imaginary mass would lead to dispersion relations characteristic for tachyons,
with propagation speed greater than $c$, which we do not accept. This leaves two options only, $\mu$ real positive or zero.

\subsection{Massive quarks case}

In the case of real positive mass the sixth-order characteristic equation $E^6 = \mid {\bf p} \mid^6 c^6 = m^6 c^{12}$ 
yields the following sixth-order differential equation after applying the quantum corespondence principle according to which 
$E \rightarrow - i \hbar \partial_t, \; \; p_k \rightarrow - i \hbar \partial_k, \; \; k = 1,2,3,$
\begin{equation}
- \hbar^6 \frac{\partial^6}{\partial t^6} = - \hbar^6 \Delta^6 c^6 + m^6 c^{12}.
\label{diffeq6new}
\end{equation} 
In order to ensure positive sixth-order derivatives on both sides, with the same separation of variables as in the massless
case, imaginary values of $\omega$ and $k$  must be used as solutions, with possibility of multiplying them, as previously,
by $-1$, $\pm j$ and $\pm j^2$. 


We shall follow the same scheme as in the massless case, arranging the exponential solutions in a $3 \times 3$ matrix. As before,
 we shall chose a Cartesian frame in which the ${\bf x}$-axis is aligned with the wave vector ${\bf k}$, so that the scalar product
${\bf k} \cdot {\bf r}$ is reduced to the expression $kx$. All combinations of $\omega$, $j \omega$ and $j^2 \omega$ with $k, \; jk$
 and $j^2 k$ are displayed in the $3 \times 3$ matrix below:
\begin{equation}
\begin{pmatrix}  e^{i(\omega t - kx)} &  e^{i(j \omega t - kx)} & e^{i(j^2 \omega t - kx)} \cr
e^{i(\omega t - j kx)} &  e^{i(j \omega t - j kx)} & e^{i(j^2 \omega t - j kx)} \cr
e^{i(\omega t - j^2 kx)} &  e^{i(j \omega t - j^2 kx)} & e^{i(j^2 \omega t - j^2 kx)} \end{pmatrix}
\label{nineexp}
\end{equation}
As before, we can easily check that the matrix (\ref{nineexp}) is not only singular, but even all its $2 \times 2$ minors have a 
vanishing determinant, This means that only six out of nine solutions displayed here are linearly independent. The choice of
six out of nine is arbitrary, but for the symmetry sake we shall make the same choice as before, suppressing the three diagonal
entries, which by themselves can form a diagonal and non-singular matrix, which we shall denote by $S_{0+}$:: 

\begin{equation}
\begin{pmatrix}  e^{i(\omega t - kx)} &  0 & 0 \cr 0 &  e^{i(j \omega t - j kx)} & 0 \cr
0 & 0 & e^{i(j^2 \omega t - j^2 kx)} \end{pmatrix}
\label{diagonalexp}
\end{equation}
Without the diagonal terms we have the following non-singular matrix of solutions:
\begin{equation}
\begin{pmatrix}  0 &  e^{i(j \omega t - kx)} & e^{i(j^2 \omega t - kx)} \cr
e^{i(\omega t - j kx)} &  0 & e^{i(j^2 \omega t - j kx)} \cr
e^{i(\omega t - j^2 kx)} &  e^{i(j \omega t - j^2 kx)} & 0 \end{pmatrix}
\label{sixexp1}
\end{equation}
Another set of solutions is readily generated by taking complex conjugates of all the entries in (\ref{sixexp1}); 

\begin{equation}
\begin{pmatrix}  0 &  e^{-i(j^2 \omega t - kx)} & e^{-i(j \omega t - kx)} \cr
e^{-i(\omega t - j^2 kx)} &  0 & e^{-i(j \omega t - j^2 kx)} \cr
e^{-i(\omega t - j kx)} &  e^{-i(j^2 \omega t -  kx)} & 0 \end{pmatrix}
\label{sixexp2}
\end{equation}

The determinants of both matrices (\ref{sixexp1}) and (\ref{sixexp2}) are equal to $2$. Let us separate them in two matrices, 
each of them having determinant $1$ (which, by the way, is rather an exceprion - usually the determinant of a sum of two matrices 
IS NOT equal to the sum of their determinants!):

\begin{equation}
S_1 = \begin{pmatrix}  0 &  e^{i(j \omega t - kx)} & 0 \cr
0 &  0 & e^{i(j^2 \omega t - j kx)} \cr
e^{i(\omega t - j^2 kx)} & 0 & 0 \end{pmatrix}
\label{threeexp1}
\end{equation}
and the complex-conjugate one:
 \begin{equation}
S_2 = \begin{pmatrix}  0 &  0 & e^{i(j^2 \omega t - kx)} \cr
e^{i(\omega t - j kx)} &  0 & 0 \cr
0 &  e^{i(j \omega t - j^2 kx)} & 0 \end{pmatrix}
\label{threeexp2}
\end{equation}
with det ($S_1$) = det ($S_2$) = $1$.
The remaining six independent solutions are obtained by changing the sign of $\omega$ while keeping everything else unchanged
(which corresponds to introducing solutions with negative frequencies, or with the reversed arrow of time).
Any linear combination of the above set od solutions will satisfy the sixth-order equation; therefore, combining solutions
with their complex conjugates, we can produce a basis of real solutions, containing sine and cosine functions multiplied
by real exponential ones. Here are two $3 \times 3$ matrices with two types of independent solutions, with cosine and sine functions;
let us denote them by $M_{1+}$ and $M_{2+}$,
the matrix $M_{1+}$ being defined as
{\small 
\begin{equation}
\begin{pmatrix} 0 &  e^{- \frac{\sqrt{3}}{2} \omega t } \cos (\frac{\omega}{2}t + kx) &  0 \cr 
0 & 0 &  e^{\frac{\sqrt{3}}{2} (\omega t +kx) } \cos (\frac{\omega}{2}t -\frac{k}{2} x) \cr
e^{-\frac{\sqrt{3}}{2} kx } \cos (\omega t + \frac{k}{2} x ) & 0 & 0 \end{pmatrix} 
\label{MatrixM1}
\end{equation} }
and the matrix $M_{2+}$ defined as
{\small 
\begin{equation}
\begin{pmatrix} 0 & 0 & e^{\frac{\sqrt{3}}{2} \omega t } \sin (\frac{\omega}{2} t + kx) \cr
e^{\frac{\sqrt{3}}{2} kx } \sin (\omega t+ \frac{k}{2} x) \cr 
0 &   e^{- \frac{\sqrt{3}}{2}(\omega t + kx)} \sin (\frac{\omega}{2} t - \frac{k}{2} x) )  & 0  \end{pmatrix}
\label{MatrixM2}
\end{equation} }
The six independent solutions displayed in the two matrices $M_{1+}$ and $M_{2+}$ should be completed by another set 
of six independent solutions which are obtained by changing the sign of $\omega$ (negative frequency solutions), or, 
eqivalently, by changing the sign of $t$, which is related to particle-antiparticle symmetry of the Dirac equation.
We can display again these solutions in two $3 \times 3$ matrices, and denote them by $M_{1-}$ and $M_{2-}$. Obviously,
these matrices are also non-singular, and contain similar exponentially damping of exponentially growing factors.

Let us show that there exist some cubic or quadratic combinations of the apparently unphysical solutions that result in
finite cosine of sine functions of $t$ and $x$, whose linear combinations can represent freely propagating waves.

Upon a closer glance at the matrices we notice that the three real exponents sum up to zero in each of them. And indeed,
taking the determinants confirms this: let us start with det ($M_{1+}$ for example. The result contains only finite trigonometric 
functions:
\begin{equation}
{\rm det} (M_{1+} ) = \cos(\frac{\omega}{2} t - kx) \; \cos (\frac{\omega}{2} t - \frac{k}{2} x) \; \cos ( \omega t + \frac{k}{2} x).
\label{detM1+}
\end{equation} 
Using the standard trigonometric formulas, we get the linearized expression:
\begin{equation}
{\rm det} (M_{1+} ) = \frac{1}{4} \; \left[ \cos (2 \omega t + kx) + \cos (\omega t + 2 kx) + \cos (\omega t - kx) + 1 \right]
\label{detM1+lin}
\end{equation}
In addition to the three running waves we get a constant term $\frac{1}{4}$ which is obviously unphysical.
But we can easily get rid of it by substracting det $(M_{1-}$ from the above expression: as we saw, it can be obtained from 
the expression (\ref{detM1+lin}) just by replacing $\omega$ by $- \omega$; therefore, we get, using the fact that $\cos \alpha = \cos(- \alpha)$:
$$ {\rm det} (M_{1+} ) - {\rm det} (M_{1-}) = \frac{1}{4} \left[ \cos (2 \omega t + kx) - \cos (2 \omega t - kx) \right] $$
\begin{equation}
+ \frac{1}{4} \left[  \cos (\omega t + 2 kx) - \cos (\omega t - 2 kx) + \cos (\omega t + kx) - \cos (\omega t - kx) \right] .
\label{M1diff}
\end{equation}

The determinants of the $M_{2+}$ and $M_{2-}$ matrices display a similar structure, with cubic expressions of sine functions in place
 of the cosine ones, but with the same arguments and the same real exponentials that cancel each other in the determinant. The 
final result is:
$$ {\rm det} (M_{2+} ) - {\rm det} (M_{2-}) = \frac{1}{4} \left[ \sin (\omega t - kx) + \sin (\omega t + kx) \right] $$
\begin{equation}
+ \frac{1}{4} \left[  \sin (\omega t - 2 kx) + \sin (\omega t + 2 kx) - \sin (2 \omega t - kx) - \sin ( 2 \omega t + kx) \right] .
\label{M2diff}
\end{equation}
Quite obviously, the same expressions appear in the third powers of matrices, which turn out to be proportional to the $3 \times 3$
unit matrix. multiplied by the corresponding determinant. We have therefore:
$$(M_{1+})^3 = {\rm det} (M_{1+}) \;   {\mbox{l\hspace{-0.55em}1}}_{3}, \; \; \; 
(M_{2+})^3 = {\rm det} (M_{2+}) \;   {\mbox{l\hspace{-0.55em}1}}_{3}, $$
\begin{equation}
(M_{1-})^3 = {\rm det} (M_{1-}) \;   {\mbox{l\hspace{-0.55em}1}}_{3}, \; \; \; 
(M_{2-})^3 = {\rm det} (M_{2-}) \;   {\mbox{l\hspace{-0.55em}1}}_{3},
\label{fourcubes}
\end{equation}
which is not surprising given the form of these off-diagonal traceless matrices. It can be easily checked that the relation (\ref{fourcubes}) 
between the determinant and the third power of any matrix of this kind hold automatically. \footnote{This is not true for the diagonal matrix
(\ref{diagonalexp}), neither for its real and imaginary parts.}


However, a serious problem remains: although the sine and cosine functions describe free planar waves, the dispersion relations between $\omega$ and ${\bf k}$
are inherited from the sixth-order characteristic equation $\omega^6 - \mid {\bf k} \mid^6 = \mu^6$, and do not reproduce the usual relativistic dispersion relations 
of the Klein-Gordon equation, where we have $\omega^2 - \mid {\bf k} \mid^2 = m^2$

It is clear that is no reason to restraint the ternary products to three solutions with identical frequency (energy) and wave vector (the 3-momentum). Let us
find out whether it is possible to obtain a running wave with relativistic second-order dispersion relation by taking a product of three solutions of the
coloured Dirac equation with three different $\omega$'s and ${\bf k}$'s, all of them satisfying the $6$-th order dispersion relation. Let us consider three
different solutions having the standard exponential form, with three different four-vectors $k^{\mu}_i, \; \; i=1,2,3$, all of them satisfying the sixth-order
dispersion relation, possibly with different masses (including massless case, too), to make the ewample as general as possible. We have explicitly:
\begin{equation}
f_1 = e^{i (\omega_1 \; t - {\bf k}_1 \cdot {\bf r})}, \; \;   f_2 = e^{i (\omega_2 \; t - {\bf k}_2 \cdot {\bf r})}, \; \;   
f_3 = e^{i (\omega_3 \; t - {\bf k}_3 \cdot {\bf r})},
\label{Threeomegas}
\end{equation}   
altogether $12$ real parameters $\omega_i, {\bf k}_l$, satisfying three $6$-th order dispersion relations
\begin{equation}
\frac{\omega_1^6}{c^6} - \mid {\bf k}_1 \mid^6 = m_1^6, \; \; \;  \frac{\omega_1^6}{c^6} - \mid {\bf k}_1 \mid^6 = m_1^6, \; \; \;  
\frac{\omega_1^6}{c^6} - \mid {\bf k}_1 \mid^6 = m_1^6,
\label{Threedispers6}
\end{equation}  
the above three constraints leaving only $9$ free parameters out of $12$.

As we have shown previously, each choice of a $4$-vector $k^{\mu}_i$ generates twelve independent exponential solutions, corresponding to different choices
of complex solutions to dispersion relations (\ref{Threedispers6}) corresponding to six complex roots of $1$:
\begin{equation}
(\pm \omega_1,  \pm j \; \omega_1,   \pm j^2 \; \omega_1), \; \;  
(\pm \omega_2,   \pm j  \omega_2,   \pm j^2 \; \omega_2), \; \;  
(\pm \omega_3,  \pm j  \omega_1,   \pm j^3 \; \omega_1),
\label{sixomegas}
\end{equation} 
and similarly
\begin{equation}
(\pm {\bf{k}}_1,  \pm j \; {\bf{k}}_1,  \pm j^2 \; {\bf{k}}_1), \; \;  
(\pm {\bf{k}}_2,  \pm j \; {\bf{k}}_2,  \pm j^2 \; {\bf{k}}_2), \; \;  
(\pm {\bf{k}}_3,  \pm j \; {\bf{k}}_3,  \pm j^2 \; {\bf{k}}_3).  
\label{sixkas}
\end{equation} 

As we know, most of the exponential solutions produced with those wave vectors, e.g. $e^{i (j^{\alpha} \omega_l t - j^{\beta} {\bf k}_l \cdot {\bf r})}$ (with $\alpha, \; \beta = 0,1, 2$)
will be damped due to the presence of imaginary parts in $j$ or $j^2$.

Now we want to answer the following question: is it possible to produce a free running wave $e^{i(\Omega t - {\bf K} \cdot {\bf r})}$ with both $\Omega$ and ${\bf K}$ real,
by taking a product of three different (damped) solutions of the colour Dirac equation? In other words, can we find linear combinations of $j^{\alpha} \omega_k $ and $j^{\beta} {\bf k}_l $ 
with vanishing maginary parts, and satisfying the usual quadratic relativistic relation $\Omega^2 - {c^2} {\bf K}^2 = M^2 $ ? And if so, how many independent solutions
can be produced in this manner?

Let us form the following combinations od $\omega_i$ and ${\bf k}_l$:

$$\Omega = j^{\alpha} \omega_1 + j^{\beta} \omega_2 + j^{\gamma} \omega_3, \; \; \; \alpha, \beta, \gamma = 0, 1, 2, $$
\begin{equation}
{\bf K} = j^{a} {\bf k}_1 + j^{b} {\bf k}_2 + j^{c} {\bf k}_3, \; \; \; a,b,c = 0,1,2.
\label{BigOmegaK}
\end{equation} 
with $\Omega$ and ${\bf K}$ satisfying reality conditions (4 independent equations):
\begin{equation}
{\it{Im}} (\Omega) = 0, \; \; \; {\it{Im}}( {\bf K} ) = {\bf 0},
\label{realityOK}
\end{equation}
(four real equations) and the relativistic invariance condition:
\begin{equation}
\Omega^2 - {c^2} {\bf K}^2 = M^2
\label{OKrelativ}
\end{equation}
altogether extra $5$ constraints. Substracted from the $9$ we had, there are only FOUR degrees of freedom left - the exact amount needed to parametrize the
space of all $4$-vectors. Thus, the system of $8$ equations imposed on $12$ real parameters defining the $12$-component solution of coloured Dirac field, 
is just enough to fix an arbitrary solution of the Klein-Gordon equation produced as a product of three independent colour Dirac solutions. Let us illustrate this
by the following example.
   
Consider three independent solutions of the colours Dirac equation, defined by three $4$-vectors with frequencies and wave vectors aligned along randomly chosen
 complex $Z_3$ generators, e.g. $(1, \; j, \; j^2)$ for $\omega_k$ and $(j, \; 1, \; j^2)$ for ${\bf k}_l$. The product of the corresponding three exponentials
is an exponentioal $e^{i (\Omega t - {\bf K} \cdot {\bf r})} $ with the following exponents:
\begin{equation}
\Omega = \omega_1 + j \omega_2 + j^2 \omega_3, \; \; \; {\bf K} = j {\bf k}_1 + {\bf k}_2 + j^2 {\bf k}_3.
\label{OmKas}
\end{equation}  
To make both $\Omega$ and ${\bf K}$ real, we must ensure that
\begin{equation}
\omega_2 = \omega_3 \; \; \; \; {\bf k}_1 = {\bf k}_3, \; \;  {\rm whence}  \; \; \Omega = \omega_1 - \omega_2, \; \; \; {\bf K} = {\bf k}_2 - {\bf k}_1.
\label{OKreal}
\end{equation} 
If $[\frac{\Omega}{c}, {\bf K} ]$ are supposed to transform as a Lorentz $4$-vector, so must do $[ \frac{\omega_l}{c}, \; {\bf k}_l ], \; m=1,2,3$
But how can it be if instead of the second-order Lorentz-invariant dispersion relation they obey the sixth-order ones?  Nevertheless, we can show that
$4$-vectors with entries multiplied by $j$ or $j^2$ undergo Lorentz transformations with appropriate coefficients. 
As a matter of fact, the $6$-th order dispersion relation is a product of three equations defining complex mass shells, more exactly, a real one
 and two mutually conjugate complex ones:
\begin{equation}
k_0^2 - {\bf k}^2 = m^2, \; \; \; k_0^2 - j {\bf k}^2 = j m^2, \; \; \; k_0^2 - j^2 {\bf k}^2 = j^2 m^2,
\label{masshells}
\end{equation}
The three representations of the same $4$-vector transform according to three equivalent representations of the same Lorentz transformation,
as shown explicitly in the Appendix II.

\subsection{Massless quarks case}

In the case when $\mu =0$, the characteristic equation yields this simple dispersion relation, valid for real or imaginary $\omega$ and ${\bf k}$ alike:
\begin{equation}
\frac{\omega^6}{c^6} = \mid {\bf k} \mid^6.
\label{dispzero}
\end{equation}
As in the massive quarks case, any particular solution gives automatically rise to six other possibilities resulting from
multiplication by sixth roots of unity, which are $\pm 1, \;  \pm j, \; \pm j^2$. It is enough to multiply $\omega$ by these six numbers;
${\bf k}$ is a three-dimensional vector, so its values cover all directions on the unit sphere, therefore there is no need to use the minus
sign, the three values related by the $Z_3$ symmetry are enough: ${\bf k}, \; j {\bf k}, \; j^2 {\bf k}.$ Taking this into account, we can
form all possible products of elementary solutions for $T(t)$ and $X (x)$ in exponential form, and display them in the following $3 \times 3$
matrices:  

\begin{equation}
\begin{pmatrix}  e^{\omega t - {\bf k}\cdot {\bf r}} & e^{\omega\,t - j {\bf k \cdot r}} & e^{\omega\,t - j^2 {\bf k \cdot r}} \cr 
e^{j \omega\,t - {\bf k \cdot r }} &  e^{j \omega t - j {\bf k}\cdot {\bf r}} & e^{j \omega\,t - j^2 {\bf k \cdot r}} \cr
e^{j^2 \omega\,t - {\bf k \cdot r}}& e^{j^2 \omega\,t - {j \bf k \cdot r}}  & e^{j^2 \omega t - j^2 {\bf k}\cdot {\bf r}} \end{pmatrix}, 
\label{twomatrices1}
\end{equation}
and a similar matrix with negative frequencies:
\begin{equation}
\begin{pmatrix} e^{- \omega t + {\bf k}\cdot {\bf r}} & e^{-\omega\,t - j {\bf k \cdot r}} & e^{-\omega\,t - j^2 {\bf k \cdot r}} \cr 
e^{-j \omega\,t- {\bf k \cdot r }} & e^{- j \omega t - j {\bf k}\cdot {\bf r}} & e^{-j \omega\,t - j^2 {\bf k \cdot r}} \cr
e^{-j^2 \omega\,t -{\bf k \cdot r}}& e^{-j^2 \omega\,t - j {\bf k \cdot r}}  & e^{- j^2 \omega t - j^2 {\bf k}\cdot {\bf r}} \end{pmatrix}
\label{twomatrices2}
\end{equation}
 
Similarly to the massive case, both matrices (\ref{twomatrices1}, \ref{twomatrices2}), have vanishing determinants; 
moreover, any of their $2 \times 2$ minors have vanishing determinants, too. 

To produce a basis of $12$ independent solutions we must suppress three items in each of the $3 \times 3$ matrices of solutions (\ref{twomatrices1}, \ref{twomatrices2}).
The most elegant way to do this is to suppress the diagonal terms. 
It is interesting to observe that each of the three diagonal items suppressed can be obtained from certain cubic combinations of the remaining ones. 
Taking into account that all $2 \times 2$ minors of the full $3 \times 3$ matrices \ref{twomatrices1}, \ref{twomatrices2} are singular, we have, for example, 
\begin{equation}
{\rm det} \; 
\begin{pmatrix}  e^{\omega t - {\bf k}\cdot {\bf r}} & e^{\omega\,t - j {\bf k \cdot r}}  \cr 
e^{j \omega\,t - {\bf k \cdot r }} &  e^{j \omega t - j {\bf k}\cdot {\bf r}}  \end{pmatrix} = 0.
\label{detminor1}
\end{equation}
But this means that 
\begin{equation}
e^{\omega t - {\bf k}\cdot {\bf r}} \; e^{j \omega t - j {\bf k}\cdot {\bf r}} -   e^{\omega\,t - j {\bf k \cdot r}} \;   
e^{j \omega\,t - {\bf k \cdot r }} = 0, 
\label{detminor2}
\end{equation}
from which we infer that 
\begin{equation}
e^{\omega t - {\bf k}\cdot {\bf r}} = e^{\omega\,t - j {\bf k \cdot r}} \;  e^{j \omega\,t - {\bf k \cdot r }} \;  e^{-j \omega t + j {\bf k}\cdot {\bf r}}.
\label{detminor3}
\end{equation}
Note that in the case when $\omega$ and ${\bf k} $ are chosen to be pure imaginary, this means that a free running wave solution can be constructed as a product
of three solutions with damping factors, provided they are taken in an appropriate way - two with positive and one with negative frequency. A similar construction
can result from another minor containing $e^{\omega t - {\bf k}\cdot {\bf r}}$, and two other ones resulting from complex conjugation, making the total number
of independent running wave solutions equal to $4$. We shall investigate this more closely in the next section.

Let us choose (among many other possible ones) of $12$-dimensional basis of independent solutions to the system (\ref{systemsix}) is
the following choice displaying a nice symmetry - and consisting in keeping the off-diagonal entries only:
\begin{equation}
\begin{pmatrix}  0 & e^{\omega\,t + j {\bf k \cdot r}} & e^{\omega\,t + j^2 {\bf k \cdot r}} \cr 
e^{j \omega\,t + {\bf k \cdot r }} & 0 & e^{j \omega\,t + j^2 {\bf k \cdot r}} \cr
e^{j^2 \omega\,t + {\bf k \cdot r}}& e^{j^2 \omega\,t + {j \bf k \cdot r}}  & 0 \end{pmatrix}, \; \; 
\begin{pmatrix}  0 & e^{\omega\,t + j {\bf k \cdot r}} & e^{\omega\,t + j^2 {\bf k \cdot r}} \cr 
e^{j \omega\,t + {\bf k \cdot r }} & 0 & e^{j \omega\,t + j^2 {\bf k \cdot r}} \cr
e^{j^2 \omega\,t + {\bf k \cdot r}}& e^{j^2 \omega\,t + {j \bf k \cdot r}}  & 0 \end{pmatrix}.
\label{offdiag}
\end{equation}
We can also produce similar matrix with six real functions, taking appropriate linear combinations as follows:
{\small
\begin{equation}
\begin{pmatrix}   0 &   e^{\omega\,t - \frac{{\bf k \cdot r}}{2}}
\, \cos ({\bf K} \cdot {\bf r}) &  e^{\omega\,t - \frac{{\bf k \cdot r}}{2}} \, \sin ({\bf K} \cdot {\bf r}) \cr 
 e^{- \frac{\omega\,t}{2}+{\bf k \cdot r }} \, \cos \Omega \, t & 0 &
e^{- \frac{\omega\,t}{2}- \frac{{\bf k \cdot r}}{2}} \, \cos (\Omega \, t  - {\bf K} \cdot {\bf r}) \cr
e^{- \frac{\omega\,t}{2} +{\bf k \cdot r}} \, \sin \Omega \, t & 
e^{- \frac{\omega\,t}{2}- \frac{ {\bf k \cdot r}}{2}} \, \sin (\Omega \, t - {\bf K} \cdot {\bf r}) & 0 
\end{pmatrix},
\label{rmatrixsix}
\end{equation} }
where $\Omega=\frac{\sqrt{3}}{2} \, \omega$ and  ${\bf K}=\frac{\sqrt{3}}{2}{\bf k}$; a similar matrix,
of course, can be produced for the alternative choice with negative $\omega$ .

\subsection{Ternary combinations producing asymptotic free waves} 
In what follows, we shall replace the scalar product of the wave vector ${\bf k}$ aith the radius-vector ${\bf r}$ by
simple expression $k_x x$, because it is always possible to choose a coordinate frame in which the the axis $Ox$ is
aligned along the vector ${\bf k}$, the two other axes perpendicular to it. For the sake of simplicity, the scalar
product will be denoted without lower index $x$ under the wave vector; $kx$ will be used instad.

The solutions displayed in (\ref{offdiag}) and (\ref{rmatrixsix}) contain damping or exploding exponential factors, 
and cannot propagate farther than a few wavelengths.
However certain quadratic or cubic products can be chosen in such a way that these nasty factors will mutually cancel
leaving a freely propagating sinusoidal wave. For example, choosing two solutions from basis different than the one
displayed in (\ref{twomatrices1}), namely 
\begin{equation}
e^{j \omega t + j^2 k x} \cdot e^{-j^2 \omega t - j kx} =
e^{(j-j^2) \; \omega t - (j^2 -j) \; kx} = e^{i (\Omega t - K x)},
\label{doublerun}
\end{equation} 
with $\Omega = \frac{\sqrt{3}}{2} \omega, \; \; \; \; K = \frac{\sqrt{3}}{2} k.$ 
An alternative choice of two elementary solutions leads to a free wave running in the opposite
direction:
$$e^{-j^2 \omega t - j^2 kx} \cdot e^{j \omega t + j kx} = e^{i ( \Omega t + K x)}.$$ 
This are the unique quadratic combinations yelding a running wave in both directions. This suggests that quark-anti-quark
states behave as Lorentz scalars.

We face different situation if we consider possibilities of producing a running wave with cubic expressions involving solutions
displayed in (\ref{rmatrixsix}) and their conjugates with negative $\omega$'s. Let us produce the following basis of real
solutions of (\ref{systemsix}):
$$ R_1 = e^{- \frac{\omega t}{2} + kx} \; \sin \Omega t, \; \; \; \; \; \; \;
 R_2 =e^{- \frac{\omega t}{2} + kx} \; \cos \Omega t, $$
$$G_1 = e^{\omega t - \frac{kx}{2}} \; \sin Kx, \; \; \; \; \; \; \; 
G_2 = e^{\omega t - \frac{kx}{2}} \; \cos Kx, $$
\begin{equation}
B_1 = e^{- \frac{\omega t}{2} - \frac{kx}{2}} \; \sin (\Omega t - Kx), \; \; \; \; \; 
B_2 = e^{- \frac{\omega t}{2} - \frac{kx}{2}} \; \cos (\Omega t - Kx).
\label{FGH12bis}
\end{equation} 

Let us display the six independent positive frequency solutions in two traceless matrices (of grade $1$, according to the $Z_3$-grading 
of $3 \times 3$ traceless matrices):
\begin{equation}
M_1 = \begin{pmatrix} 0 & R_1 & 0 \cr 0 & 0 & G_1 \cr B_1 & 0 & 0 \end{pmatrix}, \; \; \;  
M_2 = \begin{pmatrix} 0 & R_2 & 0 \cr 0 & 0 & G_2 \cr B_2 & 0 & 0 \end{pmatrix}, 
\label{M1M2matrices}
\end{equation}
Similar two matrices can be formed for the negative frequency version, replacing $\omega$ by $- \omega$, thus completing the $12$
basic solutions of our system.

We have two ways to produce ternary combinations of basic solutions: taking determinants of $M_1$ and $M_2$, or forming ternary products
of matrices and then taking the trace. Due to the fact that trace of a product of matrices does not depend on the ordering, we have
only the following independent possibilities: $M_1^3, \; M_2^2, \; M_1 M_2 M_1$ and $M_2 M_1 M_2$. These operations lead to the following results:
\begin{equation} 
{\rm det} (M_1) = R_1 G_1 B_1, \; \; \; {\rm det} (M_2) = R_2 G_2 B_2
\label{detM1M2}
\end{equation}
(note that only different letters are present in ternary products). The traces of two products $M_1 M_2 M_1$ and $M_2 M_1 M_2$ yield
the following expressions:
\begin{equation}
{\rm Tr} ( M_1 M_2 M_1) = R_1 G_2 B_1 + G_1 B_2 R_1 + B_1 R_2 G_1, 
\label{TraceM1}
\end{equation}
\begin{equation}
{\rm Tr} ( M_2 M_1 M_2) = R_2 G_1 B_2 + G_2 B_1 R_2 + B_2 R_1 G_2.
\label{TraceM2}
\end{equation}
And now, a kind of miracle occurs: in all these combinations the damping exponentials have cancelled each other, leaving only
products of bounded sinusoidal functions. 

It was shown in (\cite{RKJL2019}) that there exist only four linear combinations of ternary products of functions $R_1, G_1, B_1$
and $R_2, G_2, B_2$ defined above that lead to pure sinusoidal waves. Out of the eight cubic products without damping factors,
$$R_1 \; G_1 \; B_1, \; \; \; \; R_1 \; G_1 \; B_2, \; \; \; \; R_1 \; G_2 \; B_1, \; \; \; \; R_1 \; G_2 \; B_2; $$
\begin{equation}
R_2 \; G_1 \; B_1, \; \; \; \; R_2 \; G_1 \; B_2, \; \; \; \; R_2 \; G_2 \; B_1, \; \; \; \; R_2 \; G_2 \; B_2. 
\label{eightFGH}
\end{equation}
only two combinations with positive frequencies and two combinations with negative frequencies can be formed.
The two positive-frequency ones are:
\begin{equation}
R_1 (G_2 B_2 + G_1 B_1) + R_2 (G_2 B_1 - G_1 B_2) = \sin (2 \Omega t - 2 K x),
\label{running1b}
\end{equation}
\begin{equation}
R_2 ( G_2 B_2 + G_1 B_1) + R_1 ( G_1 B_2 - G_2 B_1) = \cos (2 \Omega t - 2 K x).
\label{running2b}
\end{equation}
Two similar running waves are produced by forming corresponding cubic combinations
of negative frequency solutions obtained by substituting $- \omega$ instead of $\omega$. We shall get then
four independent running waves, two in one direction, two in its oppowite. This circumstance suggests
that the three-quark combnations may behave like four-component standard Dirac spinors.

\section{Propagators}


Let us start with recalling the concise form of the coloured generalization of the Dirac equation written in a manifestly
$4$-dimensional form with Minkowskian space-time indices $\mu, \nu = 0,1,2,3$:
\begin{equation}
\Gamma^{\mu} p_{\mu} \; \Psi = m c \; \; {\mbox{l\hspace{-0.55em}1}}_{12} \; \Psi, \; \; \; 
{\rm with} \; \; p^0 = \frac{E}{c}, \; \; p^k = [ \; p^x, p^y, p^z \; ].
\label{Gammasecond2}
\end{equation}
with $12 \times 12$ matrices $\Gamma^{\mu} \; \; (\mu = 0, 1, 2, 3)$ defined as follows: 
\begin{equation}
\Gamma^0 =  B^{\dagger} \otimes \sigma_3 \otimes {\mbox{l\hspace{-0.55em}1}}_2, \; \; \; \; \; 
\Gamma^{k} =  Q_2 \otimes (i \sigma_2) \otimes  {\sigma}^k
\label{Gammasbig1}
\end{equation}
Our $12$-component colour Dirac equation is obviously invariant under an arbitrary similarity transformation, i.e. if we set  
\begin{equation}
\Psi' = {\cal{R}} \; \Psi, \; \; \; (\Gamma^{\mu})' = {\cal{R}} \; \Gamma^{\mu} \; {\cal{R}}^{-1} \; \; \; 
{\rm then} \; \; \; (\Gamma^{\mu})' p_{\mu} \; \Psi' = mc \; \Psi',
\label{similgamma}
\end{equation}
we get obviously  
\begin{equation}
\left[ (\Gamma^{\mu})' p_{\mu} \right]^6 = (p_0^6 - \mid {\bf p} \mid^6 ) \; {\mbox{l\hspace{-0.55em}1}}_{12}
\label{Diracprim}
\end{equation}
Following the formulae (\ref{Gammasbig}) for the colour Dirac $\Gamma^{\mu}$-matrices we see that they are neither real 
(${\bar{\Gamma}}^{\mu} \neq \Gamma^{\mu}$) nor Hermitean ($(\Gamma^{\mu})^{\dagger} \neq \Gamma^{\mu}$). 

From the colour Dirac equation (\ref{Gammasecond2}) one gets the equations for complex-conjugated ${\bar{\Psi}}$ 
and Hermitean-conjugated $\Psi^{\dagger}$: 
\begin{equation}
{\bar{\Gamma}}^{\mu} p_{\mu} \; {\bar{\Psi}} = mc \; {\bar{\Psi}}, \; \; \; \; \; \; 
p_{\mu} \Psi^{\dagger} (\Gamma^{\mu})^{\dagger} = mc \Psi^{\dagger},
\label{PsiPsibar}
\end{equation}
where ${\bar{\Psi}}$ is a column, $\Psi^{\dagger}$ is a row, ${\bar{\sigma}}_k = - \sigma_2 \sigma_k \sigma_2$, 
$\sigma_k = \sigma^k, \; \sigma_0 = \sigma^0 = {\mbox{l\hspace{-0.55em}1}}_{2}$, and 
$${\bar{\Gamma}}^0 = B \otimes \sigma_3 \otimes \; {\mbox{l\hspace{-0.55em}1}}_{2}, \; \; \; 
{\bar{\Gamma}}^k = Q_1 \otimes (i \sigma_2) \otimes {\bar{\sigma}}^k,$$
\begin{equation}
(\Gamma^0)^{\dagger} = B \otimes \sigma_3 \otimes \;  {\mbox{l\hspace{-0.55em}1}}_{2}, \; \; \;
(\Gamma^k)^{\dagger} = Q_1 \otimes \sigma_3 \otimes \sigma^k,
\label{Gammasbar}
\end{equation}
The second equation of (\ref{PsiPsibar}) can be written in terms of matrices $\Gamma^{\mu}$ if we introduce the Hermitean-adjoint 
colour Dirac spinor $\Psi^H = \Psi^{\dagger} C$ , where the $12 \times 12$-matrix $C$ satisfies the relation 
\begin{equation}
(\Gamma^{\mu})^{\dagger} C = C \Gamma^{\mu}.
\label{CGammaC}
\end{equation}
It can be also shown that neither ${\bar{\Gamma}}^{\mu}$ nor $(\Gamma^{\mu})^{\dagger}$ can be obtained via similarity transformation. 

To obtain a general solution of the colour Dirac equation one should use its Fourier transformed version, using the momentum representation: 
\begin{equation}
\left( \Gamma^{\mu} \, p_{\mu} - m \; {\mbox{l\hspace{-0.55em}1}}_{12} \right) \; {\hat{\Psi}} (p) = 0.
\label{Dirterfour}
\end{equation}
The sixth power of the matrix $\Gamma^{\mu} p_{\mu}$ is diagonal and proportional to $m^6$, so that we have 
\begin{equation}
 \left( \Gamma^{\mu} p_{\mu} \right)^6 - m^6  \; {\mbox{l\hspace{-0.55em}1}}_{12}   =
\left( p_0^6 - \mid {\bf p} \mid^6 - m^6 \right) \;  {\mbox{l\hspace{-0.55em}1}}_{12} =  0.
\label{Gammasixem}
\end{equation}
To find the inverse of the matrix $\left( \Gamma^{\mu} \, p_{\mu} - m \; {\mbox{l\hspace{-0.55em}1}}_{12} \right)$.
let us note that the sixth-order expression on the left-hand side in (\ref{Gammasixem}) can be factorized as follows: 
{\small \begin{equation}
 \left( \Gamma^{\mu} p_{\mu} \right)^6 - m^6 = \left( \left(\Gamma^{\mu} p_{\mu} \right)^2 - m^2 \right) \;
\left(  \left( \Gamma^{\mu} p_{\mu} \right)^2 - j \; m^2 \right) \; \left( \left( \Gamma^{\mu} p_{\mu} \right)^2 - j^2 \; m^2 \right).
\label{Gammafact3}
\end{equation} }
The first factor can be expressed as the product of two linear operators, one of which defines the colour Dirac equation 
 (\ref{Gammasecond2}), (\ref{Dirterfour}):
\begin{equation}
\left( \Gamma^{\mu} p_{\mu} \right)^2 - m^2 =
\left( \Gamma^{\mu} p_{\mu} - m \right) \;\left( \Gamma^{\mu} p_{\mu} + m \right) \;
\label{Gammafact4}
\end{equation}
Therefore the inverse of the Fourier transform of the linear operator defining the colour Dirac equation (\ref{Dirterfour}) 
is given by the following matrix: 
\begin{equation}
\left[ \Gamma^{\mu} p_{\mu}  - m \right]^{-1} =
\frac{\left( \Gamma^{\mu} p_{\mu} + m \right) \;
\left(  \left( \Gamma^{\mu} p_{\mu} \right)^2 - j \; m^2 \right) \;
\left( \left( \Gamma^{\mu} p_{\mu} \right)^2 - j^2 \; m^2 \right)}{\left( p_0^6 - \mid {\bf p} \mid^6 - m^6 \right)}.
\label{Dirac3inverse}
\end{equation} 
Each of the remaining five factors of the first order can be inversed in a similar way, e.g.  
\begin{equation}
\left[ \Gamma^{\mu} p_{\mu}  -j  m \right]^{-1} =
\frac{\left( \Gamma^{\mu} p_{\mu} + j m \right) \;
\left(  \left( \Gamma^{\mu} p_{\mu} \right)^2 -  m^2 \right) \;
\left( \left( \Gamma^{\mu} p_{\mu} \right)^2 - j \; m^2 \right)}{\left( p_0^6 - \mid {\bf p} \mid^6 - m^6 \right)}.
\label{Dirac3jinverse}
\end{equation} 
It is important to stress once again that the overall power counting in (\ref{Dirac3jinverse}) shows that the dominant high momentum behavior is like $p^{-1}$, 
and in the limit of $\mid {\bf p} \mid << mc$ behaves like $m^{-1}$, just like in the case of classical Dirac equation and its propagator,

This suggests the strategy to be chosen for finding the explicit form of propagators in the spacetime representation using the inverse Fourier 
transformation technique. The Fourier image of the usual Dirac field propagator can be written as 
\begin{equation}
(\gamma^{\mu} p_{\mu} - m)^{-1} = \frac{ (\gamma^{\mu} p_{\mu} + m)}{(p^{\nu}p_{\nu} - m^2)} = 
\frac{ (\gamma^{\mu} p_{\mu} + m)}{(p^2_0 - \mid {\bf p}  \mid^2 - m^2)},
\label{InverseDirac}
\end{equation}
The numerator of fraction (\ref{InverseDirac}) is a differential operator of the first order in the spacetime representation; therefore
the simplest way to obtain the Dirac propagator is to determine the spacetime propagator of the denominator, which is the inverse of a scalar
Klein-Gordon field, e.g. the Feynman propagator $\Delta_F (x^{\lambda})$, and then act on it with the differential operator $(\gamma^{\mu} p_{\mu} + m)$:
\begin{equation}
S_F (x) = (\gamma^{\mu} \partial_{\mu} + m) \; \Delta_F \; {\mbox{l\hspace{-0.55em}1}}_{4},
\label{SpaceDiracprop}
\end{equation}
where we tensorised the Feynman Klein-Gordon propagator by the unit $4 \times 4$ matrix, because the Dirac operator is $4 \times 4$
matrix valued differential operator. It is always easier to differentiate than to perform non-trivial integrations.

A serious problem remains though: the numerators contain derivatives up to the fifth order, instead of being of the first order like in the case
of the usual Dirac equation. Let us show that it is possible to find a linear combination that cancels all higher order terms, leaving a first
order operator only. To do this, let us develop the numerator of the first example, (\ref{Dirac3inverse}), into the sum of terms of different
orders of derivation. We get, after adding together ters of the same order and using the identity $j+j^2 = -1$:
$$\left( \Gamma^{\mu} p_{\mu} + m \right) \;
\left( \left( \Gamma^{\mu} p_{\mu} \right)^2 - j  m^2 \right) \;
\left( \left( \Gamma^{\mu} p_{\mu} \right)^2 - j^2  m^2 \right) = $$
\begin{equation}  
\left( \Gamma^{\mu} p_{\mu} \right)^5 + mc \left( \Gamma^{\mu} p_{\mu} \right)^4 + m^2 c^2 \left( \Gamma^{\mu} p_{\mu} \right)^3 
- m^3 c^3  \left( \Gamma^{\mu} p_{\mu} \right)^2 + m^4 c^4  \left( \Gamma^{\mu} p_{\mu} \right) + m^5c^5. 
\label{Propag5}
\end{equation}
The inverse of the next first-order factor, $(\Gamma^{\mu} p_{\mu} + m)^{-1}$, is easily obtained from the above (\ref{Propag5}) by mere change
of the sign of the mass, $m \rightarrow -m$, yielding the folllowing expression:
\begin{equation}  
\left( \Gamma^{\mu} p_{\mu} \right)^5 - mc \; \left( \Gamma^{\mu} p_{\mu} \right)^4 + m^2 c^2 \left( \Gamma^{\mu} p_{\mu} \right)^3 
+ m^3 c^3  \left( \Gamma^{\mu} p_{\mu} \right)^2 + m^4 c^4  \left( \Gamma^{\mu} p_{\mu} \right) - m^5c^5. 
\label{Propag5A}
\end{equation}
In the sum of the above two expressions the terms containing odd powers of $(mc)$ mutually cancel, while the even terms add up, and the remaining
terms are:
\begin{equation}
\left(\Gamma^{\mu} p_{\mu} - mc \right)^{-1} + \left(\Gamma^{\mu} p_{\mu} + mc \right)^{-1} =
2  \left( \Gamma^{\mu} p_{\mu} \right)^5 + 2 m^2 c^2 \left( \Gamma^{\mu} p_{\mu} \right)^3 + 2 m^4 c^4  \left( \Gamma^{\mu} p_{\mu} \right).
\label{PropthreeA}
\end{equation} 
It is easy to form two similar expressions obtained from the above by substituting $m$ by $j m$ and by $j^2 m$; we get then, after adding
pairs with $\pm jm$ and $\pm j^2m$, the following results:
\begin{equation}
\left(\Gamma^{\mu} p_{\mu} - jmc \right)^{-1} + \left(\Gamma^{\mu} p_{\mu} + j mc \right)^{-1} =
2  \left( \Gamma^{\mu} p_{\mu} \right)^5 + 2 j^2 m^2 c^2 \left( \Gamma^{\mu} p_{\mu} \right)^3 + 2 j m^4 c^4  \left( \Gamma^{\mu} p_{\mu} \right).
\label{PropthreeB}
\end{equation} 
and 
\begin{equation}
\left(\Gamma^{\mu} p_{\mu} - j^2 mc \right)^{-1} + \left(\Gamma^{\mu} p_{\mu} + j^2 mc \right)^{-1} =
2  \left( \Gamma^{\mu} p_{\mu} \right)^5 + 2 j m^2 c^2 \left( \Gamma^{\mu} p_{\mu} \right)^3 + 2 j^2 m^4 c^4  \left( \Gamma^{\mu} p_{\mu} \right).
\label{PropthreeC}
\end{equation} 
The three expressions contain all the six propagators corresponding to the six complex masses, i.e. six complex poles similar to those of the
Lee-Wick model. Let us show how we can cancel all higher derivative terms, leaving only the numerator with a single first-order differential operator, 
similarly to the usual Dirac propagator.

Let us denote the first sum (\ref{PropthreeA}) by $A$, the next sum (\ref{PropthreeB}) by $B$ and the last one (\ref{PropthreeC}) by $C$. The sum
$A+B+C$ cancels all terms leaving only one, namely $6 (\Gamma^{\mu}p_{\mu})^5)$. There are two other combinations with complex coefficients $j$ and $j^2$:
\begin{equation}
A + j B + j^2 C = 6  m^2 c^2 \left( \Gamma^{\mu} p_{\mu} \right)^3, 
\label{DiracThird}
\end{equation}
and
\begin{equation}
A + j^2 B + j C = 6  m^4 c^4 \left( \Gamma^{\mu} p_{\mu} \right).
\label{DiracFirst}
\end{equation} 
The last expression, which is a linear combination involving six terms with coefficients exhausting the entire complex representation
of the $Z_6$ cyclic symmetry group, i.e. $\pm 1, \pm j$ and $\pm j^2$, enables us to cancel all higher derivative terms, leaving
only the first order operator, and avoid the anomalies. In order to get a similar result in the case of Lee-Wick model, Anselmi and Piva (\cite{AnselmiPiva})
constructed propagators combining numerous sophisticated integration contours in the complex plane; here the result is achieved due to particular
properties of the $Z3 \times Z2$ symmetry of our equations. 

The remaining propagators can be obtained from this one by adding particular homogeneous solutions - because substracting any propagator from
another one yields a solution of the homogeneous equation.   

From now on we can follow the strategy applied in the case of the usual Dirac equation: in the first run, obtain the spacetime representation
of the inverse of the sixth-order polynomial in the space of momenta, applying the inverse Fourier transformation, then act on
it with the first-order differential operator serving as numerator in (\ref{DiracFirst})

The inverse of the six-order polynomial can be decomposed into a sum of three expressions with second-order denominators, 
divided by the common factor of order $4$ . Let us denote by $\Omega$ the sixth root of $( \mid {\bf p} \mid^6 + m^6 )$, 
\begin{equation}
\Omega = \root6\of{ \mid {\bf p} \mid^6 + m^6},
\label{BigOmega1}
\end{equation}
along with five other root values obtained via multiplication by consecutive powers of the sixth root of unity, $q = e^{\frac{2 \pi i}{6}}$.
Recalling the definition of $j$ and that $q^2 = j$, we have the identity 
\begin{equation}
(p_0^6 - \Omega^6) = (p_0^2 - \Omega^2)((p_0^2 - j \Omega^2)((p_0^2 - j^2 \Omega^2) 
\label{Decomp1}
\end{equation}
which leads to the decomposition formula 
\begin{equation}
\frac{1}{\left( p_0^6 - \mid {\bf p} \mid^6 - m^6 \right)} = \frac{1}{3 \; \Omega^4} \left[ \frac{1}{p_0^2 - \Omega^2} +
\frac{1}{j^2 p_0^2 - \Omega^2} + \frac{1}{j p_0^2 - \Omega^2} \right]
\label{InverseK}
\end{equation}
After such a substitution in (\ref{Dirac3inverse}), six $Z_6$-graded simple poles do appear in the Fourier-transform of the propagator
(\ref{InverseK}), with two real ones $\pm \Omega$ and two conjugate Lee-Wick poles $\pm j \Omega, \; \pm j^2 \Omega$. (\cite{LeeWick}, \cite{AnselmiPiva})

As long as there is a non-zero mass term, we do not encounter the infrared divergence problem at $\mid {\bf p} \mid \rightarrow 0$.
Each of the three inverses of a second-order polynomial can be in turn expressed as a sum of simple first-order poles, e.g.  
\begin{equation}
\frac{1}{p_0^2 - j \Omega^2} = \frac{j}{2 \; \Omega} \left[ \frac{1}{p_0 -j^2 \Omega} - \frac{1}{p_0 + j^2 \Omega} \right]
= \frac{j^2}{2 \; \Omega} \left[ \frac{1}{j p_0 - \Omega} - \frac{1}{j p_0 + \Omega} \right],
\label{decompK}
\end{equation}
and similarly for other terms in (\ref{InverseK}).  

In order to introduce the propagators in the coordinate space, one has to perform the contour integrals in complex energy plane. 
The inverse Fourier transformation from the $4$-momentum into the space-time dependent functions implies 
the extension of the $p_0$ component (the energy) into the complex domain. 

The first term in the decomposition (\ref{InverseK}) of the colour Dirac propagator presents two simple poles on the real line, 
while the second and the third terms display two simple poles each, located on complex straight lines $Im p_0 = j Re p_0$ 
and $Im p_0 = j^2 Re p_0$.

Such a situation occurs in the Lee-Wick model (see \cite{LeeWick}, \cite{AnselmiPiva}), which is based on the following scalar field lagrangian:

\begin{equation} 
{\cal{L}} = \frac{1}{2} \partial_{\mu} \varphi \left(1 + \frac{\Box^2}{M^2} \right) \partial^{\mu} \varphi - \frac{m^2}{2} \varphi \left(1 + \frac{\Box^2}{M^2} \right) \varphi
- \frac{\lambda}{4} \varphi^4.
\label{LWlagrangian}
\end{equation}
The lagrangian (\ref{LWlagrangian}) is manifestly relativistic-invariant, and leads to the sixth-order equations of motion. The corresponding propagator
 - or rather, its Fourier transform - displays six simple poles in the complex $p_0$ plane:

\begin{equation}
{\hat{D}} (p^2, m^2, M) = \frac{ M^4}{(p_0^2 - \mid {\bf{p}} \mid^2 - m^2)\left( (p_0^2 - \mid {\bf{p}} \mid^2)^2 + M^4 \right) }
\label{LWprop}
\end{equation}
The four extra simple poles (residues) are placed symmetrically by pairs, one pair in the upper part of the complex $p_0$ plane, another pair below the real axis.
The two usual poles situated on the real axis complete the picture, so that we have six simple poles, as it should be for a non-degenerate sixth order differential
equation.

To get the Green functions defined in the Minkowski spacetime one should perform an inverse Fourier transform, by taking the integral of ${\hat{D}}$ with respect to
four variables $(p_0, {\bf p}) $ and the factor $e^{-i p_{\mu}x^{\mu}}$, 
\begin{equation}
G(x^{\nu}, m^2, M) = \int_{-\infty}^{+ \infty} dp^0 d^3 {\bf{p}} \frac{ M^4 e^{-i p_{\mu} x^{\mu}}}{(p_0^2 - \mid {\bf{p}} \mid^2 - m^2)\left( (p_0^2 - \mid {\bf{p}} \mid^2)^2 + M^4 \right) }
\label{Greenf}
\end{equation}

The articles (\cite{AnselmiPiva}) discuss multiple integration paths in the complex $p_0$ plane, including or avoiding chosen residues in order to define various Green functions.
The unitarity of the Lee-Wick model is also claimed to be proven.

The decomposition (\ref{decompK}) of the inverse of the sixth-order invariant into six fractions contains $\Omega$, the sixth root of the sixth-order expression $(\mid {\bf p} \mid^6 + m^6)$,
 and we are unable to perform the integral effectively. Unfortunately, the entire invariant $(p_0^6-\mid {\bf p}\mid^6-m^6)$ does not admit decomposition into a product of
three (and then six) second-order (respectively, linear) expressions. The closest decomposition formula is as follows:

$$ (p_0^6 - \mid {\bf p} \mid^6 - m^6) = $$
\begin{equation}
(p_0^2 - \mid {\bf p} \mid^2 - m^2) (p_0^2 - j \mid {\bf p} \mid^2 - j^2 m^2) (p_0^2 - j^2 \mid {\bf p} \mid^2 - j m^2)
 + 3 (p_0^2 \mid {\bf p} \mid^2 m^2).
\label{decompthreeK}
\end{equation} 
The last term containing the $m^2$ factor disappears in the massless case, making a direct calculus of inverse Fourier transform feasible.

\subsection{The massless case - using convolution}

In the massless case, the operator equation whose Green's function we want to evaluate, reduces to
$$ \left[ \frac{1}{c^6} \frac{\partial^6}{\partial t^6} - \left( \frac{\partial^2}{\partial x^2} + \frac{\partial^2}{\partial y^2}
+ \frac{\partial^2}{\partial z^2} \right)^3 \right] \; G (t, {\bf r}) = \delta^4 (x) = \delta(ct) \delta(x) \delta(y) \delta(z).$$
Using the Fourier transformation method, we can write:
\begin{equation} 
\left[ \frac{\omega^6}{c^6} - \mid {\bf p} \mid^6 \right] \; {\hat{G}} (p_0, {\bf p}) = 1, \; \; \; \; {\rm where} \; \; p_0 = \frac{\omega}{c},
\label{FourierGreen}
\end{equation}
from which we get 
\begin{equation}
{\hat{G}} (p_0, {\bf p} ) =  \frac{1}{p_0^6 - \mid {\bf p} \mid^6 } + \Phi (p_0, {\bf p} ),
\label{FGreen2}
\end{equation}
where $\Phi (p_0, {\bf p} )$ is a solution of the homogeneous equation, 
\begin{equation}
\left[ p_0^6 - \mid {\bf p} \mid^6 \right] \; \Phi (p_0, {\bf p} ) = 0 \; \; \; {\rightarrow} \; \; \; \Phi (p_0, {\bf p} ) 
= \delta (p_0^6 - \mid {\bf p} \mid^6 ).
\end{equation} 
Remembering that in the case of the Klein-Gordon equation, the homogeneous solution in the momentum space, $\delta (p_0^2 - \mid {\bf p} \mid^2)$,
could be decomposed into a sum of two delta-functions corresponding to the factorization of the quadratic relativistic invariant into two linearly independent
factors, which gives, neglecting numerical factors,
\begin{equation}
\delta (p_0^2 - \mid {\bf p} \mid^2) \simeq \delta (p_0 - \mid {\bf p} \mid) + \delta (p_0 + \mid {\bf p} \mid ).
\label{Twodeltas}
\end{equation}
corresponding to retarded or advanced solutions.

The sixth-order polynomial $p_0^6 - \mid {\bf p} \mid^6 $ can be split into the product of three second-order factors as follows:
\begin{equation}
p_0^6 - \mid {\bf p} \mid^6 = (p_0^2 - \mid {\bf p} \mid^2) \; (p_0^2 - j \mid {\bf p} \mid^2) \; (p_0^2 - j^2  \mid {\bf p} \mid^2),
\label{kthree}
\end{equation}
each of which can be split again into two linear factors, thus giving SIX different homogeneous solutions, which can be arranged in three pairs,
the first one as in (\ref{kthree}) above, the next two pairs given by
\begin{equation}
 \delta (p_0 - j \mid {\bf p} \mid) + \delta (p_0 + j \mid {\bf p} \mid ), \; {\rm and} \; \;  \delta (p_0 - j^2 \mid {\bf p} \mid) + \delta (p_0 + j^2 \mid {\bf p} \mid ),
\label{sixdeltas}
\end{equation}
and we should remember another set of six delta-functions with negative $p_0$, corresponding to negative frequencies, all in all twelve independent solutions
of the homogeneous system. 

One could be puzzled by the appearance of delta-functions with complex arguments, which do not seem to be well defined. However, they correspond to complex mass shells
(in the massive case), or the $Z_3$ complex generalizations of the light-cone. They correspond to twelve exponential solutions obtained by direct computation, with frequencies 
taking values $\pm \omega, \; \pm j \omega, \; \pm j^2 \omega$ combined with complex wave vectors ${\bf k}, \; j {\bf k}$ and $j^2 {\bf k}$, displayed in matrices 
(\ref{sixexp2}, \ref{threeexp1}) and (\ref{threeexp2}).

We have now two different ways to treat the problem. Either we decompose the inverse of (\ref{kthree}) into simple fractions, with six first-order poles,
as follows:
\begin{equation}
\frac{1}{p_0^6 - \mid {\bf p} \mid^6} = \frac{1}{3 \mid {\bf p} \mid^4} \; \left[ \frac{1}{p_0^2 - \mid {\bf p} \mid^2} +  \frac{j}{p_0^2 - j \mid {\bf p} \mid^2} +
\frac{j^2}{p_0^2 - j^2 \mid {\bf p} \mid^2} \right],
\label{threemid}
\end{equation}
(with each of the three second-order inverses split later into a sum of two first-order inverses, displaying explicitly six different simple poles
at $\pm \mid {\bf p} \mid, \; \; \pm j \; \mid {\bf p} \mid$ and $\pm \; j^2 \mid {\bf p} \mid$. But we get a fourth-order singular factor
in front of these, $\mid {\bf p} \mid^{-4}$, which may make the Fourier integration problematic. 

Or we can keep the decomposition into product of three factors, each of them an inverse of a second-order expression, which will remove the problem of the
$\mid {\bf p} \mid^{-4}$ formfactor in front of the sum (\ref{threemid}) and the research of its inverse Fourier image.

What remains now is the integration over $d {\bf p}$ and $d p_0$. As usual, the integral over $dp_0$ is taken first, and evaluated
by extension to the complex domain. The first factor $G_1$ has two poles on the real line, $p_0 = \pm \mid {\bf p} \mid$, and is evaluated
as a principal value. The final result is the well known Green's function of the d'Alembertian. The remaining two factors
are integrated over $d p_0$ even more easily, because their poles are found off the real axis, as shown in the following figure:

\begin{figure}[hbt]
\centering
\includegraphics[width=5cm, height=3.8cm]{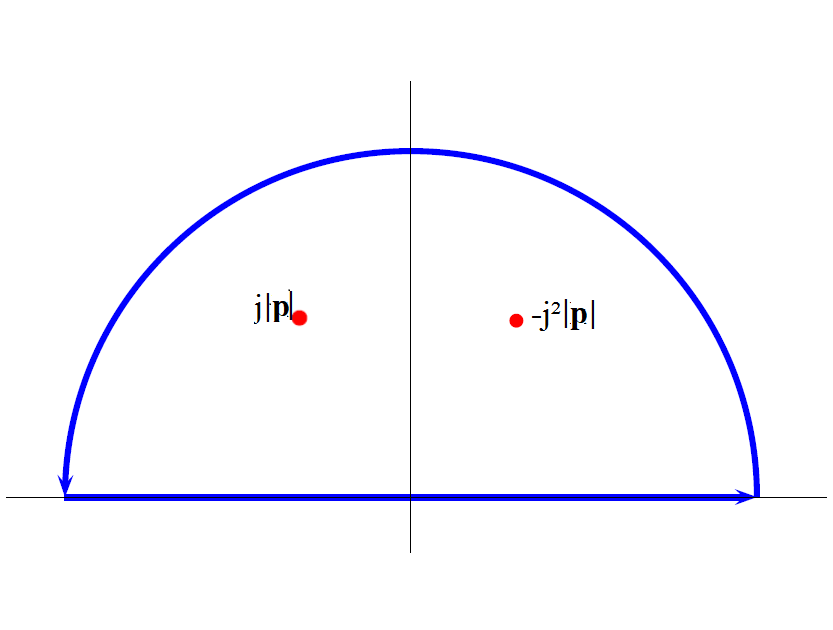}
\hskip 0.2cm
\includegraphics[width=5cm, height=3.8cm]{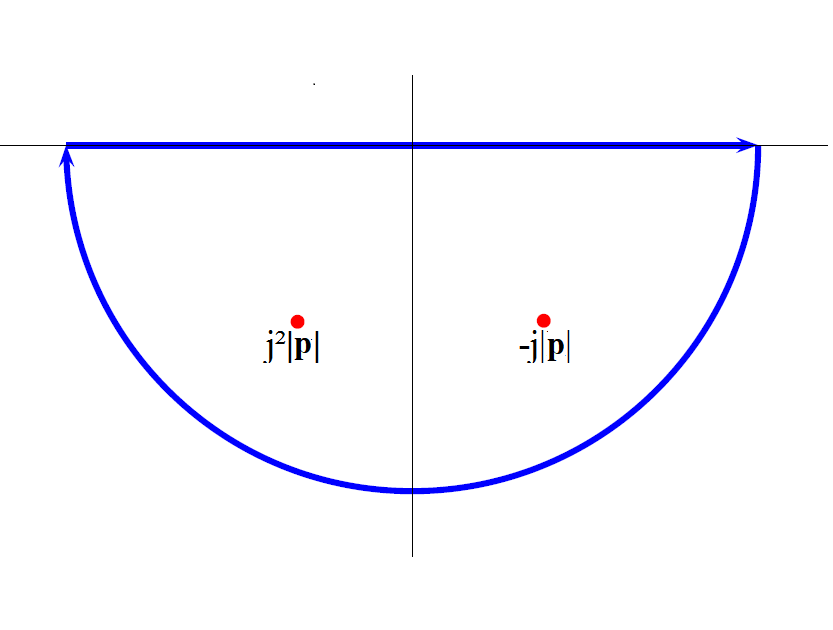}
\caption{{\small Left: The upper contour, containing the poles at $j \mid {\bf p} \mid$ and $- j^2 \; \mid {\bf p} \mid$, for $t < 0$; 
Right: The lower contour, containing the poles at $ -j \mid {\bf p} \mid$ and $j^2 \; \mid {\bf p} \mid$, for $t > 0$ . }}
\label{fig:Contours}
\end{figure}

The Green functions corresponding to the remaining two factors, with two alternative couples of poles, can be obtained by similar integration
performed along the real axis, without being forced to avoid the real poles: the four remaining ones are found over or under the real axis,
as shown in the figures below:
\begin{figure}[hbt]
\centering
\includegraphics[width=4.7cm, height=3.8cm]{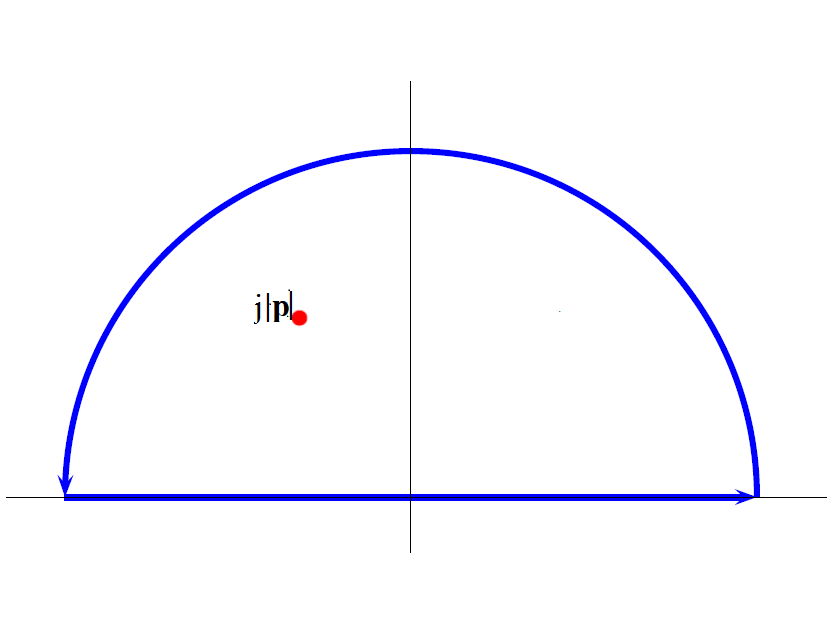}
\hskip 0.4cm
\includegraphics[width=4.8cm, height=3.9cm]{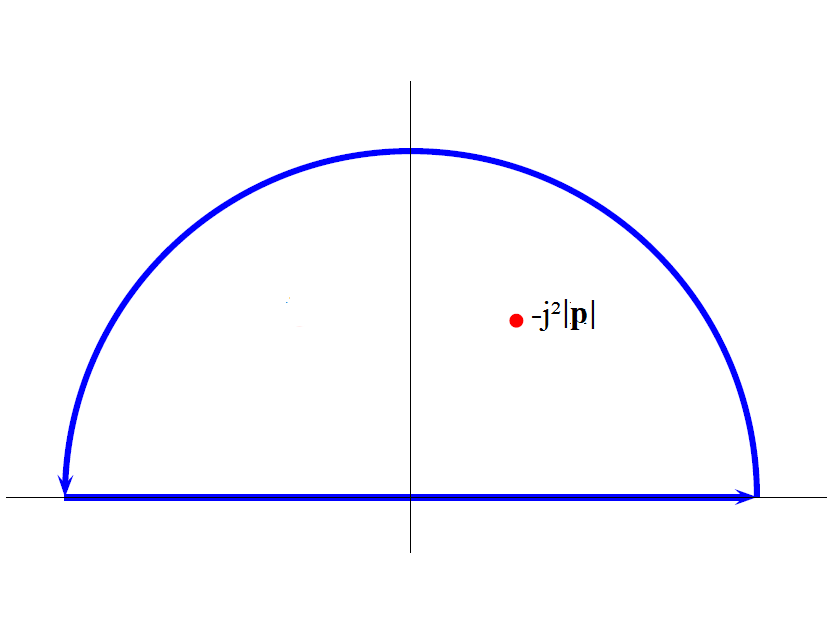}
\caption{{\small Two upper contours, containing the poles at $j \mid {\bf p} \mid$ and $- j^2 \; \mid {\bf p} \mid$, for $t > 0$;  }}
\label{fig:Contours_ppup}
\end{figure}
\begin{figure}[hbt]
\centering
\includegraphics[width=4.7cm, height=3.8cm]{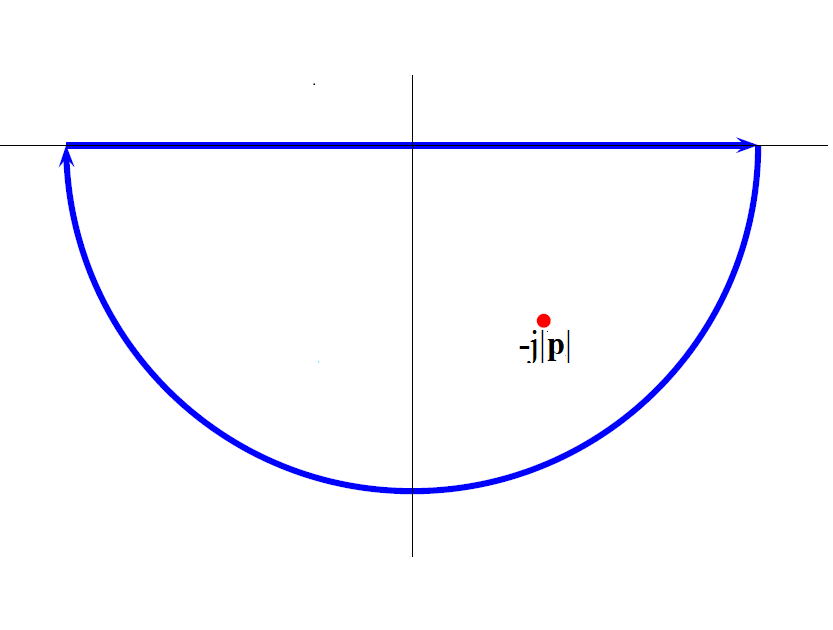}
\hskip 0.4cm
\includegraphics[width=4.8cm, height=3.9cm]{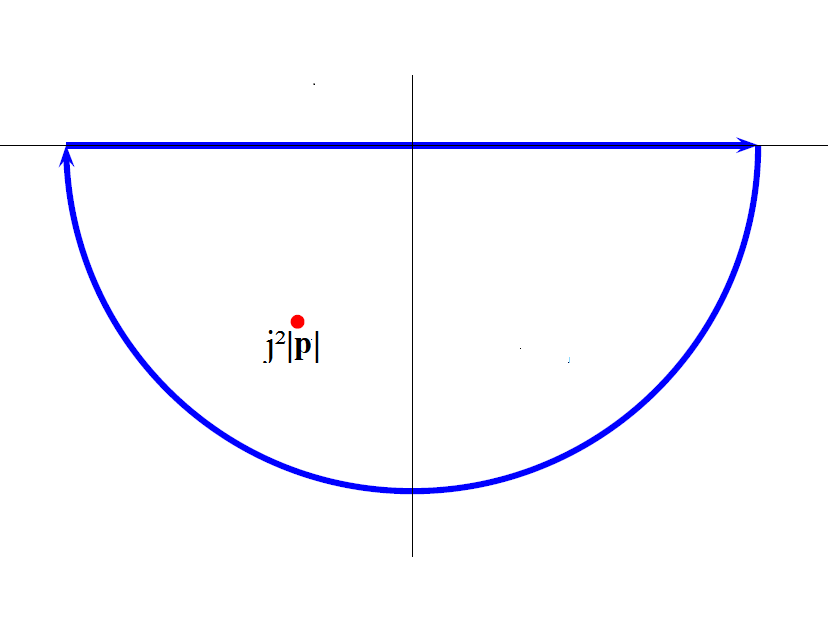}
\caption{{\small Two lower contours with poles at $-j \mid {\bf p} \mid$ and at $j^2 \mid {\bf p} \mid$, for $t<0$.}}
\label{fig:Contours_ppdown}
\end{figure}
The two poles in the upper half of the complex plane, $j \mid {\bf p} \mid$ and $-j^2 \mid {\bf p} \mid $, belong to two different terms 
of the decomposition into three fractions, the second and the third one, and so do the two poles $ -j \mid {\bf p} \mid$ and $j^2 \; \mid {\bf p} \mid$
situated in the lower half of the complex plane. The results of integration in the complex $p_0$-plane should be appropriately combined in order
to get simple and elegant and real form. Here are the two resulting combinations: the upper plane integrals performed for $t > 0$, the lower plane
integrals performed for $t<0$ are

\begin{equation}
2 \pi Y(t) \; \frac{1}{\mid {\bf p} \mid} \;  e^{- \frac{\sqrt{3}}{2} \omega t} \; \sin \frac{\omega t}{2} \; \; \; \;  {\rm and} \; \; 
- 2 \pi Y(- t) \; \frac{1}{\mid {\bf p} \mid} \; e^{\frac{\sqrt{3}}{2} \omega t} \; \sin \frac{\omega t}{2},
\label{Twodamped}
\end{equation} 
which combine the results of complex integrations of two inverses, $(p_0^2 - j \mid {\bf p}\mid^2)^{-1}$ and $(p_0^2 - j^2 \mid {\bf p}\mid^2)^{-1}$.  
  

\subsection{The massless case - with decomposition in simple fractions}

Performing a triple convolution with Heaviside and Dirac delta distributions depending on complex arguments poses some problems.
Let us try an alternative approach instead: decomposing the Fourier image of the inverse operator in the massless case into
a sum of simpler operators multiplied by a common formfactor, and try to find the original of the $\mid p \mid^4$ formfactor
separately, and after that perform the convolution.

We have to find out the general form of the Green's function whose Fourier image is given by
$${\hat{G}} (p) = \frac{1}{p_0^6 - \mid {\bf p} \mid^6} = \frac{1}{3 \mid {\bf p} \mid^4}  \left[ \frac{1}{p_0^2 - \mid {\bf p} \mid^2}
+ \frac{j}{p_0^2 - j \mid {\bf p} \mid^2} +  \frac{j^2}{p_0^2 - j^2 \mid {\bf p} \mid^2} \right].$$
The presence of the factor $\mid {\bf p} \mid^{-4}$, strongly diverging at $\mid {\bf p} \mid \rightarrow 0$ (an infrared divergence), seems
to pose a problem, as compared with the mild factor $\mid {\bf p} \mid^{-1}$ in a similar decomposition in the usual Klein-Gordon equation
of the second order. However, we shall show that it can be circumvened using a well known regularization technique. Before applying it to the
common formfactor $\mid {\bf p} \mid^{-4}$, let us find the spacetime original of the sum of three fractions:
\begin{equation}
\left[ \frac{1}{p_0^2 - \mid{\bf p}\mid^2} +   \frac{j}{p_0^2 - j \mid{\bf p}\mid^2} + \frac{1}{p_0^2 -j^2 \mid{\bf p}\mid^2} \right]
\label{Threefrac}
\end{equation}

What remains now is to find the spacetime original of the formfactor $\frac{1}{ \mid {\bf p} \mid^4 }$. 
Let us make an ``educated guess'' that the result of the inverse Fourier integral transforming 
${\hat{G}}(p)$ into $G(r)$ is a function of $r$ of the form
\begin{equation}
V(r) = \alpha \; r^{\beta}, 
\label{V(r)}
\end{equation}
with $\alpha$ and $\beta$ unknown real constants. In order to avoid divergence of the integral, let us multiply this function by a Yukawa-like
potential $e^{-\lambda r}$, with some real positive parameter $\lambda$, which will be made tend to zero after integral is performed.
The volume element in thee dimensions is given by $r^2 \sin \theta dr \; d \theta d \varphi$; integration over $ d \varphi$ from $0$ to $2 \pi$
will yield the factor $2 \pi$; also, as usual, integration with respect to the polar angle $\theta$ from $0$ to $\pi$ can be replaced by
integration over $u=\cos \theta$ because $\sin \theta \; d \theta = - d (\cos \theta)$. After this standard procedure, the integral to be taken is given by:
\begin{equation}
2 \pi \alpha \int_0^{\infty} \; r^2 dr \int_{-1}^1 du \; r^{\beta} \; e^{-i k r u} e^{- \lambda r}  
\label{integralu}
\end{equation}
The integral over $d u$ gives:
\begin{equation}
\int_{-1}^1 \;  e^{-i k r u} \; du = \frac{e^{-ikr} - e^{ikr}}{-ikr} = \frac{2 \sin (kr)}{kr},
\label{sinpr}
\end{equation}
What is left now is the integration over $dr$:
\begin{equation}
2 \pi \alpha \; \int_0^{\infty} \; r^{2+ \beta -1} e^{- \lambda r} \frac{ 2 \sin (k r)}{k} dr.
\label{Integral_kr}
\end{equation}
A short Mathematica program (see the Appendix II) gives the answer according to three possible choices of the exponent $\beta$ (up to a numerical factor in front of the result):
$$ \beta = -1: \; \; \; \; \; \frac{\alpha}{k^2 + \lambda^2} \; \; \;  ({\rm Coulomb \; potential \; when} \; \lambda \rightarrow 0,) $$
\begin{equation}
\beta = 0: \; \; \; \; \; \; \frac{\lambda}{(\lambda^2 + k^2)^2} \; \; \; ({\rm going \; to \; zero \; with} \; \lambda \rightarrow 0,
\label{betazero}
\end{equation}
$$\beta = 1: \; \; \; \; \frac{2 (k^2 - 3 \lambda^2) }{(\lambda^2 + k^2)^3 } \; \; \; \; ({\rm linear \; potential} \; V \simeq r. $$
As we see, the original of the formfactor $\mid {\bf p} \mid^{-4}$ in its most general form is a superposition of two effective potentials,
$V_1 \simeq r^{-1}$ and $V_2 \simeq r$, the form is often claimed for the effective interaction between the quarks 
(see, e.g. \cite{Neuberger}, \cite{Takahashi}).

In order to get the propagator in the space-time representation, we should perform the convolution of one of these functions tensorised with $\delta(ct)$
(keeping in mind that the convolution concerns all four space-time variables) with {\it retarded} propagators obtained above, all of them containing
the Heaviside function $Y(t)$ as common factor, because a convolution is well defined only for distributions with compact support, or with supports
bounded from below. The result of convolution with respect to the time variable of $(\alpha(r) \otimes \delta(ct)$ with any distribution depending on $[ct, x,y,z]$ 
will just reproduce the dependence on $ct$ without any modification, $\delta(ct)$ acting as unit element in the convolution algebra; what remains then is the
triple convolution with respect to space variables $(x,y,z)$, and leaving the Heaviside function $Y(ct)$ as a common factor in front of the convolution:
\begin{equation}
G(x^{\mu}) \simeq Y(ct) \left( r \right) * \left( \frac{\delta(ct - r)}{r} +  j \frac{\delta(ct - jr)}{r} + j^2 \frac{\delta(ct - j^2 r)}{r} \right)   
\label{Greenfull}
\end{equation}

\section{Interaction with gauge fields}

\subsection{Lagrangian and minimal interaction}

The lagrangian serving as the integrand for the variational principle in $4$-dimensions leading to the generalized colour
Dirac equation can be easily introduced following the same prescription that led to the usual Dirac's lagrangian
\begin{equation}
{\cal{L}} = i {\bar{\psi}} \gamma^{\mu} \partial_{\mu} \psi  + m {\bar{\psi}} \psi,  
\label{Dirlagr}
\end{equation}
where ${\bar{\psi}}$ is the Dirac conjugate of $\psi$, ${\bar{\psi}} = \psi^{\dagger} \gamma^o$, and $\psi^{\dagger}$ is the hermitian
conjugate of the Dirac spinor $\psi$.

In our case the Dirac conjugate of the generalized $12$-component spinor should be defined as follows.
\begin{equation}
{\bar{\Psi}} = \Psi^{\dagger} B \otimes \sigma_3 \otimes \; {\mbox{l\hspace{-0.55em}1}}_{2}
\label{BigDirconjug}
\end{equation}
now the lagrangian cen be written in the following concise form:
\begin{equation}
{\cal{L}}_{color} = i \left( {\bar{\psi}} \Gamma^{\mu} \partial_{\mu} \Psi \right) + {\bar{\Psi}} \mu c \Psi.
\label{LagrDirCol}
\end{equation}
Written explicitly, it shows how the original form of the system is recovered under the variational principle:
$$\Psi^{\dagger} B \otimes \sigma_3 \otimes \;  {\mbox{l\hspace{-0.55em}1}}_{2} \; \left[ B^{\dagger} \otimes \sigma_3 \otimes \; {\mbox{l\hspace{-0.55em}1}}_{2}
- Q_2 \otimes (i \sigma_2) \otimes {\boldsymbol{\sigma}}\cdot {\bf p}  \right] \; \Psi = $$
\begin{equation}
\Psi^{\dagger} p_0 \Psi - \psi^{\dagger} Q_3 \otimes \sigma_1 \otimes {\boldsymbol{\sigma}}\cdot {\bf p} \Psi,
\label{Lagrexplicit}
\end{equation}
In this form, which ensures the energy ($p_0$ is diagonal) positivity, the original system of equations (\ref{systemsix}) is clearly visible again.

The minimal interaction with gauge fields is introduced as usual, by replacing space-time derivationa $\partial_{\mu}$ by their gauge-covariant
counterparts. In the usual Dirac equation the gauge potential is that of the electromagnetic field, the coupling constant being
the elementary electric charge $e$: 
$$\partial_{\mu} \rightarrow \partial_{\mu} - e A_{\mu}$$
In our case the gauge potential should take values in the $12 \ times 12$ matrice algebra, and more precisely, in the same product representation
which may be symbolized as $Z_3 \times Z_2 \times Z_2$ scheme, i.e. the tensor product of a $3 \times 3$ matrix with a $2 \times 2$ matrix and
yet another $2 \times 2$ matrix. This combination gives space to accomodating two gauge fields, the $SU(3)$ algebra valued colour field mediating
strong interactions, and the electromagnetic field, which should couple to the usual Dirac degrees of freedom, i.e. the four Lorentzian spinorial 
degredes of freedom carried by the $4 \times 4$ factor. The resulting form of the combined strong (gluons) and electromagnetic (photons) fields'
gauge potential is given by the following substitution:

\begin{equation}
{\cal{A}} = \lambda_a \; B^a_{\mu} (x^{\nu}) \otimes {\mbox{l\hspace{-0.55em}1}}_{2} \otimes {\mbox{l\hspace{-0.55em}1}}_{2} + 
A_{\mu} (x^{\nu}) {\mbox{l\hspace{-0.55em}1}}_{3} \otimes {\mbox{l\hspace{-0.55em}1}}_{2}
\otimes {\mbox{l\hspace{-0.55em}1}}_{2}.
\label{Agluem}
\end{equation}
where $\lambda_a, \; \; a, b = 1,2,...8$ are the Gell-Mann $3 \times 3$ matrices spanning the $SU(3)$ Lie algebra. 
It is easy to check that the field tensor defined as ${\cal{F}}_{\mu \nu} =  \partial_{\mu} {\cal{A}}_{\nu} - 
\partial_{\nu} {\cal{A}}_{\mu} + [ {\cal{A}}_{\mu}, {\cal{A}}_{\nu} ]$
yields two separate tensors aligned along different subspaces of $12 \times 12$ matrices:
\begin{equation}
\partial_{\mu} {\cal{A}}_{\nu} - \partial_{\nu} {\cal{A}}_{\mu} + [ {\cal{A}}_{\mu}, {\cal{A}}_{\nu} ] = 
G^a_{\mu \nu} \; \lambda_a \otimes {\mbox{l\hspace{-0.55em}1}}_{2} \otimes {\mbox{l\hspace{-0.55em}1}}_{2} 
+ F_{\mu \nu} \; {\mbox{l\hspace{-0.55em}1}}_{3} \otimes {\mbox{l\hspace{-0.55em}1}}_{2}
\label{Gmunu}
\end{equation}
where 
\begin{equation}
G^a_{\mu \nu} = \partial_{\mu} B^a_{\nu} - \partial_{\nu} B^a_{\mu} + f^a_{cd} B^c_{\mu} B^d_{\nu}, \; \; \; F_{\mu \nu} = \partial_{\mu} A_{\nu} - \partial_{\nu} A_{\mu} ,
\label{GFfields}
\end{equation}
with $f^a_{cd}$ being the structure constants of the $SU(3)$ Lie algebra, defined by the commutators $[ \lambda_c, \lambda_d] = f^a_{cd} \lambda_a.$

\subsection{Adding the weak interactions}

Apparently, there remains no place for the weak interactions described by the $SU(2)$ generated gauge fields. 

It was shown in (\cite{RKJL2021}) that in order to close the commutation relations of the generalized $Z_3$-covering of the Lorentz algebra acting on the generalized Dirac spinors,
it is necessary to introduce the so-called ``Lorentz doublets'', i.e. pairs of coloured Dirac spinors, spanning a new $2$-dimensional representation space, and leading to
$24$-component generalized coloured spinors and $24 \times 24$ matrix operators acting on them.  

In order to take into account the structure of Lorentz doublets corresponding to the isospin content,
 the generalized potential should take its values in $24 \times 24$-matrices resulting from tensorization of the $12 \times 12$ matrices with 
$2 \times 2$ generators of the $SU(2)$ isospin algebra. This leads to the following expression of the most general gauge potential:
\begin{equation}
{\cal{A}}_{\mu} = 
\sigma_k A^{k}_{\mu} \otimes {\mbox{l\hspace{-0.55em}1}}_{2} \otimes {\mbox{l\hspace{-0.55em}1}}_{2} \otimes {\mbox{l\hspace{-0.55em}1}}_{2}  
+{\mbox{l\hspace{-0.55em}1}}_{2} \otimes \lambda_a B^{a}_{\mu} \otimes \; {\mbox{l\hspace{-0.55em}1}}_{2} \otimes {\mbox{l\hspace{-0.55em}1}}_{2} 
+ {\mbox{l\hspace{-0.55em}1}}_{2} \otimes {\mbox{l\hspace{-0.55em}1}}_{3} \otimes {\mbox{l\hspace{-0.55em}1}}_{2} \otimes A_{\mu} \; {\mbox{l\hspace{-0.55em}1}}_{2} 
\label{Biggaugepot}
\end{equation}
The gauge invariant field tensor is defined as usual,
\begin{equation}
{\cal{F}}_{\mu \nu} =
 \partial_{\mu} {\cal{A}}_{\nu} - \partial_{\nu} {\cal{A}}_{\mu} 
+ \left[ {\cal{A}}_{\mu} , {\cal{A}}_{\nu} \right]
\label{Biggaugefield}
\end{equation}
It is easy to show that the gauge field Lagrangian defined below:
\begin{equation}
{\cal{L}}_{gauge} = \frac{1}{6} \; {\rm Tr} \left( {\cal{F}}^{\mu \nu} \; {\cal{F}}_{\mu \nu} \right]
\label{Lgauge}
\end{equation}
splits into three independent parts, describing weak, strong, and electromagnetic interactions:
\begin{equation}
\frac{1}{6} {\rm Tr} \left( {\cal{F}}^{\mu \nu} \; {\cal{F}}_{\mu \nu} \right) 
= \frac{1}{4} F_k^{\mu \nu} F^k_{\mu \nu} +  \frac{1}{4}G_a^{\mu \nu} G^a_{\mu \nu} + \frac{1}{4} F^{\mu \nu} F_{\mu \nu},
\label{ThreeF}
\end{equation}
with $i, k = 1,2,3,$ \; $a, b = 1,2...8.$, and where
\begin{equation}
F^k_{\mu \nu} = \partial_{\mu} A^k_{\nu} - \partial_{\nu} A^k_{\mu} + \epsilon^{k}_{\; \; \; lm} A^l_{\mu} A^m_{\nu}, \; \; 
G^a_{\mu \nu} = \partial_{\mu} B^a_{\nu} - \partial_{\nu} B^a_{\mu} + f^a_{\; \; \; bc} B^b_{\mu} B^c_{\nu}, 
\end{equation}
are, respectively, the weak and strong gauge field tensors, $\epsilon^k_{\; \; \; lm}$ the $SU(2)$ Lie algebra structure constants, 
and $f^a_{\; \; \; bc}$ those of the $SU(3)$ algebra.

Finally, $F_{\mu \nu} = \partial_{\mu} A_{\nu} - \partial_{\nu} A_{\mu}$ is the Maxwell electromagnetic field tensor.
Thus the full gauge field content of the Standard Model is reproduced here.

\section{Concluding remarks}

The new approach to quark colour dynamics proposed in the present paper is based on the idea that all continuous symmetries take their source
at the most fundamental ones, revealed by discrete symmetries described by the simplest finite groups, the cyclic groups $Z_2$, $Z_3$, 
their product $Z_2 \times Z_3 = Z_6$ and eventually $S_3$ (full permutation group of three objects). 

The next step, consisting in construction of a genuine field theory with non-commuting operators replacing wave functions, will require
a generalization of the anti-commutation relations and the Pauli exclusion principle leading to the Fermi-Dirac statistics. A $Z_3$-graded
generalization of Pauli's principle was discussed in (\cite{Kerner2017}), where cubic $Z_3$-antisymmetric relations introduced
as early as in $1991$ (\cite{Kerner1991}) were implemented. Explaining ``in a nutshell'', the $N$ generators spanning an associative algebra 
satisfy the following cubic relations:
\begin{equation}
\theta^A \theta^B \theta^C = j \, \theta^B \theta^C \theta^A = j^2 \, \theta^C \theta^A \theta^B,
\label{ternary1}
\end{equation}
$j = e^{2 i \pi/3}$, the primitive root of $1$. 
We have $1+j+j^2 = 0$ \; \; and $\bar{j} = j^2$.

A similar set of {\it conjugate} generators is added, ${\bar{\theta}}^{\dot{A}}$, $\dot{A}, \dot{B},... = 1,2,...,N$, 
satisfying similar condition with $j^2$ replacing $j$:
\begin{equation}
{\bar{\theta}}^{\dot{A}} {\bar{\theta}}^{\dot{B}} {\bar{\theta}}^{\dot{C}} = 
j^2 \, {\bar{\theta}}^{\dot{B}} {\bar{\theta}}^{\dot{C}} {\bar{\theta}}^{\dot{A}} 
= j \, {\bar{\theta}}^{\dot{C}} {\bar{\theta}}^{\dot{A}} {\bar{\theta}}^{\dot{B}},
\label{ternary2}
\end{equation}
Finally, we complete the constitutive relations by the following $Z_3$-commutation between $\theta^A$ and ${\bar{\theta}}^{\dot{B}}$:
\begin{equation}
\theta^A {\bar{\theta}}^{\dot{B}} = -j {\bar{\theta}}^{\dot{B}} \theta^A, \; \; \;  {\bar{\theta}}^{\dot{B}} \theta^A = -j^2  \theta^A {\bar{\theta}}^{\dot{B}}.
\label{jthetathetabar}
\end{equation}
As a result, the squares of generators do not vanish, forming independent algebraic elements, but third powers of any generator identically vanish:
$$\theta^A \theta^A \neq 0,  \; \; \theta^A \theta^A \theta^A = 0.$$ 
When represented by operators acting on a Hilbert space of states, the constitutive relations (\ref{ternary1}, \ref{ternary2} and \ref{jthetathetabar})
can be interpreted as a $Z_3$ graded ternary generalization of Pauli's exclusion principle.

We endow it with a natural $Z_3$ grading, considering the generators $\theta^A$ as grade $1$ elements, their conjugates ${\bar{\theta}}^{\dot{B}}$ 
being of grade $2$. The $Z_3$ grades add up modulo $3$, so that both highest (cubic) order monomials $\theta^A \theta^B \theta^C$ and 
${\bar{\theta}}^{\dot{A}} {\bar{\theta}}^{\dot{B}} {\bar{\theta}}^{\dot{C}} $ display the $Z_3$ grade $0$.

Combined with the associativity, these cubic relations impose finite dimension on the
algebra generated by the $Z_3$ graded generators. As a matter of fact, cubic expressions are the
highest order that does not vanish; all monomials of order higher than $3$ identically vanish: 

\begin{equation}
\theta^A \theta^B \theta^C \theta^D = 0.
\label{quartic2}
\end{equation}
It was shown in (\cite{Kerner2017}) that in the lowest non-trivial case of two generators, $A, B = 1,2$ the generators $\theta^A$ and ${\bar{\theta}}^{\dot{B}}$ transform under
a $Z_3$ covering of the $SL(2, {\bf C})$ group, with $2 \times 2$ matrices $U^A_B$ whose determinants are $1, j$ or $j^2$. There are only two couples of 
independent third-order monomials, 
\begin{equation}
\phi^{1} = \theta^1 \theta^2 \theta^1, \; \; \; \phi^2 = \theta^2 \theta^1 \theta^2 \; \; \; {\rm and} \; \; \; 
{\bar{\chi}}^{\dot{1}} = {\bar{\theta}}^{\dot{1}} {\bar{\theta}}^{\dot{2}} {\bar{\theta}}^{\dot{1}}, \; \; \; 
{\bar{\chi}}^{\dot{2}} = {\bar{\theta}}^{\dot{2}} {\bar{\theta}}^{\dot{1}} {\bar{\theta}}^{\dot{2}}.
\label{phichi}
\end{equation}
When the generators $\theta^A$ and ${\bar{\theta}}^{\dot{B}}$ transform under the action of matrices $U^A_B$ with determinant $j$ and respectively
by ${\bar{U}}^{\dot{C}}_{\dot{D}}$ with determinant $j^2$, the combinations $\phi{\alpha}$ and ${\bar{\chi}}^{\dot{\beta}}$ transform under the
action of two inequivalent representations of $SL(2, {\bf C})$, like two Pauli spinors composing a Dirac spinor. Therefore, if $\theta^A$ represent
quarks, their cubic products generate behave like ordinary fermions.

A problem that could be tackled next is modelling mesons as quark-antiquark pairs, and solving a quantum two-body problem in the center of mass
coordinate system, with gauge potential $V \simeq \alpha r + \beta r^{-1}$ using the Foldy-Wouthuysen transformation. A rather clueless and
naive tentative in that direction was made in (\cite{Kerner1975}), with magnetic monopole potential for the $SU(2)$ gauge field. At that time
Quantum Chromodynamics was still in the elaboration stage and not universally accepted as nowadays. 

A challenge to be addressed in the forthcoming research is the construction of a $Z_3$-symmetric generalization of spin-statistic theorem 
for coloured spinors. The $Z_3$ anti-commutation relations should be related to non vanishing values of color and isospin, while commuting
fields would display $Z_3$ grade $0$. Quark and anti-quark fields should display respectively $Z_3$ grades $1$ and $2$, while lepton fields
sould be by definition colorless, i.e. $Z_3$-commutative (but keeping the $Z_2$ grade characteristic for fermions). 

Like the fundamental difference between fermions and bosons, here we might expect an essential difference between quarks endowed with non-zero
isospin and colour, and leptons deprived of these attributes.

\newpage

\vskip 0.1cm
\indent
\hskip 0.5cm
{\bf {\large Appendix I: The $SU(3)$ nonion representation }}
\vskip 0.3cm
\indent
The three operators appearing in the generalized Dirac's equation (\ref{systemsix}) can be expressed in terms of tensor products of matrices 
of lower dimensions. Let us introduce two following $3 \times 3$ matrices:
{\small \begin{equation}
B = \begin{pmatrix} 1 & 0 & 0 \cr 0 & j & 0 \cr 0 & 0 & j^2 \end{pmatrix} \; \; {\rm and} \; \; 
Q_3 = \begin{pmatrix} 0 & 1 & 0 \cr 0 & 0 & 1 \cr 1 & 0 & 0 \end{pmatrix}
\label{BQmatrices1}
\end{equation} }
The two $3 \times 3$ matrices $Q_3$ and $B$ are particular solutions of matrix equation defining a $3 \times 3$ matrix
representation of third root of unity matrix. They generate,
via all their products and powers, the $U(3)$ Lie group algebra, which becomes the $SU(3)$ algebra after removing the unit matrix.

The standard $3 \times 3$ matrix basis of ternary Clifford algebra (first considered in XIX-th century by Cayley 
and Sylvester, who called its elements  ``{\it nonions}'' ) looks as follows: 
{\small \begin{equation}
Q_1 = \begin{pmatrix}  0 & 1 & 0 \cr 0 & 0 & j \cr j^2 & 0 & 0 \end{pmatrix}, \; 
Q_2 = \begin{pmatrix}  0 & 1 & 0 \cr 0 & 0 & j^2 \cr j & 0 & 0 \end{pmatrix}, \; 
Q_3 = \begin{pmatrix}  0 & 1 & 0 \cr 0 & 0 & 1 \cr 1 & 0 & 0 \end{pmatrix}, \;  
\label{threeQ}
\end{equation} 
\begin{equation}
Q^{\dagger}_1 = \begin{pmatrix}  0 & 0 & j \cr 1 & 0 & 0 \cr 0 & j^2 & 0 \end{pmatrix}, \; 
Q^{\dagger}_2 = \begin{pmatrix}  0 & 0 & j^2 \cr 1 & 0 & 0 \cr 0 & j & 0 \end{pmatrix}, \; 
Q^{\dagger}_3 = \begin{pmatrix}  0 & 0 & 1 \cr 1 & 0 & 0 \cr 0 & 1 & 0 \end{pmatrix}, \; 
\label{threeQdagger}
\end{equation} } 
where $j$ is the third primitive root of unity, 
\begin{equation} j = e^{\frac{2 \pi i}{3}}, \; \; j^2 = e^{\frac{4 \pi i}{3}}, \; \; 1+j+j^2 = 0.
\label{Jot}
\end{equation}
and ${\cal{M}}^{\dagger}$ denotes the Hermitean conjugate of matrix ${\cal{M}}$. We see that all matrices $Q$ and $Q^{\dagger}$ are non-Hermitean. 
We complete the basis of $3 \times 3$ traceless matrices by adding to $Q$ and $Q^{\dagger}$ the following two linearly independent {\it diagonal} matrices: 
\begin{equation}
B = \begin{pmatrix}  1 & 0 & 0 \cr 0 & j & 0 \cr 0 & 0 & j^2 \end{pmatrix}, \; \; \; \; B^{\dagger} = 
\begin{pmatrix}  1 & 0 & 0 \cr 0 & j^2 & 0 \cr 0 & 0 & j \end{pmatrix}.
\label{twoBmatrices}
\end{equation}
The eight matrices $Q_a, \; Q^{\dagger}_b, \; B $ and $B^{\dagger}$ span a basis of the $SU(3)$ Lie algebra, with $B, \; B^{\dagger}$ its Cartan subalgebra.
The commonly used basis of traceless Gell-Mann matrices $\lambda_{\alpha}, \; \; \alpha = 1, 2, ...8$ can be obtained by linear combinations of $Q_a, \; Q^{\dagger}_b, \; B, \; B^{\dagger}$
$$\lambda_1 = \frac{1}{3} \; \left( Q_1 + Q_2 + Q_3 + Q^{\dagger}_1 + Q^{\dagger}_2 + Q^{\dagger}_3 \right),$$
$$\lambda_2 = \frac{i}{3} \; \left( Q_1 + Q_2 + Q_3 - Q^{\dagger}_1 - Q^{\dagger}_2 - Q^{\dagger}_3 \right),$$
\begin{equation}
\lambda_4 = \frac{1}{3} \; \left(j Q_1 + j^2 Q_2 + Q_3 + j^2 Q^{\dagger}_1 + j Q^{\dagger}_2 + Q^{\dagger}_3 \right),
\label{GMlambdas}
\end{equation}
$$\lambda_5 = \frac{i}{3} \; \left( j Q_1 + j^2 Q_2 + Q_3 - j^2 Q^{\dagger}_1 - j Q^{\dagger}_2 - Q^{\dagger}_3 \right),$$
$$\lambda_6 = \frac{1}{3} \; \left( j^2 Q_1 + j Q_2 + Q_3 + j Q^{\dagger}_1 +j^2 Q^{\dagger}_2 + Q^{\dagger}_3 \right),$$
$$\lambda_7 = \frac{i}{3} \; \left(j Q_1 + j^2 Q_2 + Q_3 -j^2 Q^{\dagger}_1 -j  Q^{\dagger}_2 - Q^{\dagger}_3 \right),$$

The multiplication table of matrices $Q_a, \; Q^{\dagger}_b, \; B$ and $B^{\dagger}$:
{\small
\begin{center}
\begin{tabular}{|c|c|c|c|c|c|c|c|c|}
\hline
\raisebox{0mm}[6mm][2mm]{ \, \ \ } & $Q_1$ & $Q_2$ & $Q_3$ & $Q^{\dagger}_1$ & $ Q^{\dagger}_2$ & $Q^{\dagger}_3$ & $B$ & $B^{\dagger}$ \cr
\hline\hline
\raisebox{0mm}[6mm][2mm]{ $Q_1$ } & $Q_1^{\dagger}$ & $j^2\; Q^{\dagger}_3$ & $j\;Q^{\dagger}_2$ & $ {\mbox{l\hspace{-0.55em}1}}_3 $ & $ B^{\dagger}$ & $  B$ & $ j\;Q_2$ & $ j^2 \; Q_3$ \cr
\hline
\raisebox{0mm}[6mm][2mm]{ $Q_2$ } & $j\; Q^{\dagger}_3$ & $ Q_2^{\dagger} $ & $ j^2\;Q^{\dagger}_1$ & $ B$ & $ {\mbox{l\hspace{-0.55em}1}}_3 $ & $B^{\dagger} $ & $j\; Q_3$ & $ j^2\; Q_1$ \cr
\hline
\raisebox{0mm}[6mm][2mm]{ $Q_3$ } & $j^2 Q^{\dagger}_2$ & $ j\;Q^{\dagger}_1$ & $ Q_3^{\dagger} $ & $B^{\dagger}$ & $ B $ & $ {\mbox{l\hspace{-0.55em}1}}_3 $ & $j \; Q_1$ & $ j^2 \; Q_2$ \cr
\hline
\raisebox{0mm}[6mm][2mm]{ $Q^{\dagger}_1$} & ${\mbox{l\hspace{-0.55em}1}}_3 $ & $j^2\; B $ & $j\; B^{\dagger} $ & $ Q_1 $ & $j^2\; Q_3$ & $j \; Q_2$ & $ Q^{\dagger}_3$ & $ Q^{\dagger}_2$ \cr
\hline
\raisebox{0mm}[6mm][2mm]{ $Q^{\dagger}_2$ } & $j\; B^{\dagger} $ & ${\mbox{l\hspace{-0.55em}1}}_3 $ & $ j^2\; B$ & $j \; Q_3$ & $Q_2 $ & $ j^2\; Q_1$ & $ Q^{\dagger}_1$ & $ Q^{\dagger}_3$ \cr
\hline
\raisebox{0mm}[6mm][2mm]{ $Q^{\dagger}_3$ } & $ j^2\; B$ & $j \; B^{\dagger}$ & ${\mbox{l\hspace{-0.55em}1}}_3 $ & $j^2\; Q_2$ & $j \; Q_1$ & $ Q_3 $ & $Q^{\dagger}_2$ & $Q^{\dagger}_1$ \cr
\hline
\raisebox{0mm}[6mm][2mm]{ $B$ } & $ Q_2$ &$ Q_3$ & $ Q_1$ & $j \; Q^{\dagger}_3$ & $j \; Q^{\dagger}_1$ & $ j\; Q^{\dagger}_2$ & $ B^{\dagger} $& $ {\mbox{l\hspace{-0.55em}1}}_3 $ \cr
\hline
\raisebox{0mm}[6mm][2mm]{ $B^{\dagger}$ } & $Q_3$ & $ Q_1$ & $ Q_2$ & $j^2\; Q^{\dagger}_2$ & $ j^2 \; Q^{\dagger}_3$ & $j^2 \; Q^{\dagger}_1$ & $ {\mbox{l\hspace{-0.55em}1}}_3 $ & $B$ \cr
\hline \hline
\end{tabular} 
\end{center} }
\vskip 0.2cm
{\centerline{Table 2. The multiplication table of the set of eight $3 \times 3$ traceless matrices }}
\vskip 0.3cm
\indent
As can be easily seen, the new matrices generated by multiplication of the $8$ still do not form a group. Besides the initial
set $Q_a, Q^{\dagger}_a, B$ and $B^{\dagger}$ their products produce the new matrices obtained by multiplication of the above
by $j$ and $j^2$, and the unit matrix $ {\mbox{l\hspace{-0.55em}1}}_3 $. However, the matrices $j B$ and $j^2 B^{\dagger}$ are missing,
as well as the matrices $j \;  {\mbox{l\hspace{-0.55em}1}}_3 $ and $j^2 \; {\mbox{l\hspace{-0.55em}1}}_3 $
They do appear after iteration, as products of some of the new matrices contained in the Table 2, e.g. 
$(j \; Q_1)\cdot (j \; Q_1^{\dagger} = j^2 {\mbox{l\hspace{-0.55em}1}}_3 $, and so on. Notice that the multiplication by $j$ or $j^2$
keeps the fundamental $SU(3)$ property valid: if $Q_a^{\dagger} = Q_a^{-1}$, similarly $(j \; Q_a)^{\dagger} = j^2 \; Q_a^{\dagger} = (j \; Q_a)^{-1}$,
and their determinants are also equal to $1$, because $j^3 = 1$ and $(j^2)^3=1$ as well.

The full set of $27$ elements forms a finite group, which is isomorphic with the tensor product of $9$ nonion matrices by the $Z_3$ cylic group.
This group is obviously a finite subgroup of the fundamental representation of the Lie group $SU(3)$.

The six matrices $Q_k$ and $Q_j^{\dagger}, \; i, j = 1,2,3$  are endowed with natural $\mathbb Z_3$-grading 
\begin{equation}
{\rm grade} (Q_i) = 1, \; \; \; {\rm grade} (Q^{\dagger}_k) = 2,
\label{gradeQ}
\end{equation}
The totally symmetric combination is proportional to the $3 \times 3$ identity matrix $I = {\mbox{l\hspace{-0.55em}1}}_3 $:  
\begin{equation}
 Q_a Q_b Q_c + Q_b Q_c Q_a + Q_c Q_a Q_b = 3\,\eta_{abc} \,  \;  {\mbox{l\hspace{-0.55em}1}}_3, \; \; \; a,b,... = 1,2,3.
\label{anticom}
\end{equation}
with $\eta_{abc}$ given by the following non-zero components 
$$\eta_{111} = \eta_{222} = \eta_{333} = 1, \; \; \eta_{123} = \eta_{231} = \eta_{312} = j^2, $$
\begin{equation}
\eta_{213} = \eta_{321} = \eta_{132} = j
\label{defeta}
\end{equation}
and all other components vanishing. The above relation can be used as definition of {\it ternary Clifford algebra}. 

The Hermitean conjugates $Q_{\dot{a}}^{\dagger} := {\bar{Q}}_a^T$ 
of matrices $Q_a$, which we shall endow with dotted indices ${\dot{a}}, {\dot{b}},...=1,2,3$, satisfy similar relations: 
\begin{equation}
Q_a^2 = Q_{\dot{a}}^{\dagger}
\label{defqbar}
\end{equation}
as well as the identities conjugate to the ones in (\ref{anticom}) 
 \begin{equation}
 Q^{\dagger}_{\dot{a}} Q^{\dagger}_{\dot{b}} Q^{\dagger}_{\dot{c}} + 
Q^{\dagger}_{\dot{b}} Q^{\dagger}_{\dot{c}} Q^{\dagger}_{\dot{a}} + Q^{\dagger}_{\dot{c}} Q^{\dagger}_{\dot{a}} Q^{\dagger}_{\dot{b}}
 = 3\,\eta_{{\dot{a}}{\dot{b}}{\dot{c}}} \, \; {\mbox{l\hspace{-0.55em}1}}_3,  
{\rm with} \; \;  \eta_{{\dot{a}}{\dot{b}}{\dot{c}}} = {\bar{\eta}}_{cba}.
\label{anticomdot}
\end{equation}
\vskip 0.3cm
\indent
\hskip 0.5cm
{\bf {\large Appendix II: The $Z_3$ complex Lorentz transformations }}
\vskip 0.3cm
\indent
We shall restrain our considerations to the case when the wave vector ${\bf k}$ is aligned aling the $x$-axis, reducing our example
to two dimensions, $k_0, k_x$. A Lorentz transformation induced by the change of Galielean frames with relative velocity $V$ (also
aligned along the $x$-axis) is represented by a $2 \times 2$ matrix
\begin{equation}
\begin{pmatrix} k_0^{'} \cr k_x^{'} \end{pmatrix} =  \begin{pmatrix} \cosh \; u & \sinh u \cr \sinh u & \cosh u \end{pmatrix} 
\begin{pmatrix} k_0 \cr k_x \end{pmatrix}  \; \; \rightarrow \; \; k_0^{' 2} - k_x^{'2} = k_0^2 -  k_x^2, 
\label{Lorentz2real}
\end{equation}
where we put as usual $\tanh u = V/c$ so that
\begin{equation}
\sinh u = \frac{V}{c \sqrt{ 1 - V^2/c^2}}, \; \; \; \; 
\cosh u = \frac{1}{\sqrt{ 1 - V^2/c^2}},
\label{Tanhu}
\end{equation}
The Lorentz transformation (\ref{Lorentz2real}) keeps invariant the quadratic expression $k_0^2 - k_x^2$. But so do the following two 
conjugate complex unimodular $2 \times 2$ matrices, representing the same Lorentz transformation on complex mass shells: 
\begin{equation}
 \begin{pmatrix} \cosh u & j \; \sinh u \cr j^2 \; \sinh u & \cosh u \end{pmatrix}, \; \; \; {\rm and} \; \; \;
 \begin{pmatrix} \cosh u & j^2 \; \sinh u \cr j \; \sinh u & \cosh u \end{pmatrix}, 
\label{ThreeLorentz}
\end{equation}
Both types of matrices provide a representation of the Lorentz group, each of them acting separately on the respective complex mass shells.
It is easy to see that:
\begin{equation}
\begin{pmatrix} k_0^{'} \cr j k_x^{'} \end{pmatrix} =  \begin{pmatrix} \cosh \; u & j^2 \sinh u \cr j \sinh u & \cosh u \end{pmatrix} 
\begin{pmatrix} k_0 \cr j k_x \end{pmatrix}, \; \; \rightarrow \; \; k_0^{'2} - j^2 k_x^{'2} = k_0^2 - j^2 k_x^2,  
\label{Lorentzcomplex1}
\end{equation}
and 
\begin{equation}
\begin{pmatrix} k_0^{'} \cr j^2 k_x^{'} \end{pmatrix} =  \begin{pmatrix} \cosh \; u & j \sinh u \cr j^2 \sinh u & \cosh u \end{pmatrix} 
\begin{pmatrix} k_0 \cr j^2 k_x \end{pmatrix}  \; \; \rightarrow \; \; k_0^{'2} - j k_x^{'2} = k_0^2 - j k_x^2, 
\label{Lorentzcomplex2}
\end{equation}
The generalization for arbitrary $4$-vectors with four non-vanishing components is rather straightforward, and can be found in previously published papers.
 
Obviously, if a Lorentz transformation is applied simultaneously to three replicas of the same $4$-vector $[k_0, \; {\bf k}]$, leaving invariant three
 corresponding quadratic combinations $[k_0^2 - {\bf k}^2], \; [k_0^2 -j {\bf k}^2]$ and $[k_0^2 - j^2 {\bf k}]$, then their product will produce
the sixth-order invariant $k_0^6 - \mid {\bf k} \mid^6$, making it also Lorentz-invariant - in a sense. 

It is also worth mentioning that due to the linear character of Lorentz transformations, any linear combination of two $4$-vectors keeps its form
after the transformation. In particular, if we impose the reality condition on the sum of three $4$-vectors belonging to three separate complex
mass chells, its image after a Lorentz transformation will remaibn real, too.

\vskip 0.3cm
\indent
\hskip 0.5cm
{\bf {\large Appendix III: {\it Mathematica} program of Green function's }}
\vskip 0.1cm
\indent
\hskip 3.3cm
{\bf {\large regularization }}
\vskip 0.3cm
\indent

(*The Fourier transform of potential $V(r)=r^{\beta}$*)

(*$V(q)=(2pi)*int_0^{\infty} r^2 dr int_{-1}^1$ of $V(r) e^(-i*q*r*u)*e^(-l*r)$*)

(*The regularisation factor: $e^(-lambda*r), \; lambda \rightarrow 0$ at the end of the calculus*)

Assumptions = ${l > 0, q > 0, r > 0}$;

(*Integral over angles*)

Integrate[Exp[-I*q*r*u], {u, -1, 1}]
$\beta = -1$;
Int=2 Pi*Integrate[ $r^(2 + \beta)$ 2 Sin[q*r]/(q*r)*Exp[-lambda*r], {r, 0, Infinity}];

Print[Int /. {lambda $\rightarrow$ 0}]

$\beta = 0;$

Int=2 Pi*Integrate[   $r^(2 + \beta)$ 2 Sin[q*r]/(q*r)*Exp[-lambda*r], {r, 0, Infinity}];

Print[Int /. {lambda $\rightarrow$ 0}]
$\beta = 1;$

Int=2 Pi*Integrate[  $ r^(2 + \beta)$ 2 Sin[q*r]/(q*r)*Exp[-lambda*r], {r, 0, Infinity}];

Print[Int/. {lambda $\rightarrow$ 0}]

\vskip 0.4cm
\hskip 0.7cm
{\bf Acknowledgements}
\vskip 0.4cm
We are grateful to Hadrien Kurkjian and Karol Penson for precious help with programming Fourier integration using the {\it Mathematica} facility.
One of the authors (RK) expresses his thanks to Michel Dubois-Violette and Jean-Bernard Zuber for numerous discussions and useful advices. Constructive
critical remarks by Michelangelo Mangano and John Ellis are gratefully acknowledged.

JL has been supported by the project UMO-2022/45/B/ST2/01067 from the Polish National Science Center (NCN).

\end{document}